\documentclass[a4paper,11pt]{article}
\usepackage{jheppub} % for details on the use of the package, please see the JINST-author-manual

\usepackage{physics}
\usepackage{longtable}
\usepackage{float}
\usepackage{amsthm}
\usepackage{amssymb}
\usepackage{hyperref}
\numberwithin{equation}{section}

\newtheorem{theorem}{Theorem}

\def\T{{\sf{T}}}
\def\C{\mathbb{C}}
\def\Z{\mathbb{Z}}
\def\Q{\mathbb{Q}}
\def\R{\mathbb{R}}

\def\P{\mathbb{P}}

\def\Aut{\operatorname{Aut}}

\def\Hirz[#1]{\mathbbm{F}_{#1}}
\def\o[#1]{\overline{#1}}

\def\prim{\text{prim}}
\def\Aut{\text{Aut}}
\def\ker{\mbox{ker}}

\author[a]{Andreas P. Braun,}
\author[a]{Hugo Fortin, }
\author[b]{Daniel  Lopez~Garcia}
\author[c]{Roberto Villaflor~Loyola}

\affiliation[a]{Department of Mathematical and Computing Sciences,\\ 
Durham University Upper Mountjoy Campus, \\
Stockton Rd, Durham DH1 3LE, UK\\}

\affiliation[b]{Instituto de Matem\'atica e Estat\'istica,\\ 
Universidade de S\~ao Paulo, \\
Rua do Mat\~ao 1010, S\~ao Paulo 05508-090, Brazil\\} 

\affiliation[c]{Facultad de Matemáticas,\\ 
Pontificia Universidad Católica de Chile, Campus San Joaquín, \\
Avenida Vicuña Mackenna 4860, Santiago, Chile\\} 

\emailAdd{andreas.braun@durham.ac.uk}
\emailAdd{hugo.fortin@durham.ac.uk}
\emailAdd{dflopezga@ime.usp.br}
\emailAdd{roberto.villaflor@mat.uc.cl}

\title{More on $G$-flux and General Hodge Cycles on the Fermat Sextic}

\abstract{ We study M-Theory solutions with $G$-flux on the Fermat sextic Calabi-Yau fourfold, focussing on the relationship between the number of stabilized complex structure moduli and the tadpole contribution of the flux. We use two alternative approaches to define the fluxes: algebraic cycles and (appropriately quantized) Griffiths residues. In both cases, we collect evidence for the non-existence of solutions which stabilize all moduli and stay within the tadpole bound.}

\begin{document}

\maketitle

\section{Introduction}

It is by now a classic result that fluxed IIB orientifolds, or more generally F-Theory compactifications with $G_4$ flux, lead to vast numbers of vacua \cite{Denef:2004ze,Taylor:2015xtz}. Models of this type, see \cite{Dasgupta:1999ss} for early examples, constitute the largest fraction 
of known solutions to string theory with four non-compact dimensions that can be studied with some level of technical control, as the geometry stays Calabi-Yau up to warping \cite{Becker:1996gj,Giddings:2001yu}. 
F-Theory models with flux can be described as a limit of M-Theory compactifications on Calabi-Yau fourfolds, where the effect of a flux $G$ on the geometry 
is captured by the superpotentials
\cite{Gukov:1999ya,Haack:2001jz}
\begin{equation}\label{eq:GKV_W}
\begin{aligned}
W& = \int_X \Omega \wedge G \\
\tilde{W} & = \int_X J \wedge J \wedge G\, .
\end{aligned}
\end{equation}
The stationary points are characterized by $D_I W = D_k \tilde{W} = 0$, where $I=1,\ldots, h^{3,1}(X)$ and $k=1,\ldots, h^{1,1}(X)$, which implies
\begin{equation}
\begin{aligned}
\ast G &=  G  \\
 G \wedge J &= 0  
\end{aligned}\, ,
\end{equation}
so that $G \in H^{2,2}(X) \oplus H^{4,0}(X)$. For a given choice of a primitive cohomology class of the flux, the above conditions for a vacuum can be read as 
constraints on the $h^{3,1}$ complex structure moduli of $X$. As the number of equations equals $h^{3,1}$, a `generic' choice of flux will constrain 
the stationary points of the superpotential to be points. This can also be expressed as the matrix 
\begin{equation}\label{eq:GIJmatrix}
\rho_{IJ} := D_I D_J W
\end{equation}
having maximal rank at typical minima. Supersymmetric Minkowski minima are found for stationary points of $W$ and $\tilde{W}$ which furthermore 
obey $W=\tilde{W}=0$, which implies that
\begin{equation}
\begin{aligned}
 G &\in H^{2,2}(X) \\
 G \wedge J &= 0  
\end{aligned}\, ,
\end{equation}
i.e. such fluxes are given by primitive Hodge classes. 

What makes the study of flux vacua in F/M-theory subtle is the fact that fluxes need to obey the quantization condition \cite{Witten:1996md}
\begin{equation}\label{eq:fluxquant}
 G + \frac{c_2(X)}{2} \in  H^4(X,\Z) \, ,
\end{equation}
and are furthermore constrained by cancellation of M2-brane charge to satisfy the tadpole constraint
\begin{equation}\label{eq:tadpole}
\frac{\chi(X)}{24} - \frac{1}{2} \int_X G \wedge G = N_{M2}  \, ,
\end{equation}
where $\chi(X)$ is the Euler characteristic of $X$ and $N_{M2}$ is the number of $M2$ branes filling the non-compact directions. As shown in \cite{Klemm:1996ts}, the condition \eqref{eq:fluxquant} guarantees that $N_{M2}$ is always an integer.

As the inclusion of anti M2-branes is expected to lead to instabilities triggering brane/flux annihilation, we are hence led to demanding that
\begin{equation}\label{eq:tadpole_ineq}
\frac{\chi(X)}{24} \geq \frac{1}{2} \int_X G \wedge G \, . 
\end{equation}
For $G$ primitive and self-dual, the right hand side is positive definite, so that the `length' of the flux vector is bounded by the Euler characteristic of $X$. For a given fourfold, this can be used to argue for the finiteness of flux vacua \cite{Bakker:2021uqw,Grimm:2023lrf}. 

The main motivation for the present work is to study the interplay between the number of stabilized complex structure moduli, i.e. the rank of $\rho_{IJ}$, \eqref{eq:GIJmatrix}, the tadpole constraint \eqref{eq:tadpole}, and the quantization condition \eqref{eq:fluxquant}. The classic 
counts of flux vacua estimates the number of flux solutions by approximating the number of lattice points by the volume of the ellipsoid defined by \eqref{eq:tadpole_ineq}, which is sensible if the distance of lattice points is small compared to the radii. 

What is more severely affected is the argument that $\rho_{IJ}$ is generically of full rank as the word `generic' looses its meaning for quantized fluxes. Although 
we can always rescale $G$ to come arbitrarily close to a properly quantized flux, 
the required rescaling can become very large, so that the tadpole constraint \eqref{eq:tadpole_ineq} is violated.

It is not a priori clear how restrictive the tadpole constraint really is, and what may characterize points in moduli space where all complex structure moduli may be stabilized within this bound. It was observed in \cite{Collinucci:2008pf,Braun:2011zm,Braun:2020jrx} that stabilizing complex structure moduli by the superpotential \eqref{eq:GKV_W} in F-Theory can be surprisingly expensive with regards to the contribution of fluxes to the tadpole. This observation was called the `tadpole problem' in \cite{Bena:2020xrh}, where it was conjectured that there is, at least asymptotically, a linear relationship between the number of stabilized moduli and the tadpole contribution. There are concrete examples of F-Theory models where all complex structure moduli are
stabilized within the tadpole bound \cite{Denef:2005mm,CaboBizet:2014ovf,Honma:2017uzn}, but these examples all have in common that they only have few complex structure moduli, and furthermore carry elliptic fibrations that have Weierstrass models with singularities at complex codimension $1$ in the base. From the perspective of F-Theory, this implies the exciting possibility that the points in moduli space selected by fluxes may be distinguished by the appearance of extra massless gauge bosons. 

Such a relationship was recently demonstrated for a specific class of flux solutions on $K3 \times K3$ in \cite{Braun:2023pzd}. There exists a class of fluxes
\cite{Aspinwall:2005ad} that stabilize all of the complex structure moduli resulting in pairs of `attractive' \cite{Moore:1998pn} K3 surfaces. Using the tadpole bound then leads to a finite number of solutions which have been classified in \cite{Aspinwall:2005ad,Braun:2014ola}. These can be all be lifted to F-Theory solutions by specifying an elliptic fibration as described in \cite{Braun:2013yya,Braun:2014ola}. While there are attractive K3 surfaces with elliptic fibrations that do not have reducible fibres, \cite{Braun:2023pzd} showed that the complete list of solutions within the tadpole bound is not of this type, so that the tadpole bound enforces non-abelian gauge symmetry. This result is particularly interesting in that it shows the opposite of what genericity arguments would support: as enhanced non-abelian gauge symmetries appear at loci of often considerable codimension in complex structure moduli space in F-Theory, one might conclude that they are greatly disfavoured when using the continuum approximation of the flux \cite{Braun:2014xka}. 

The present work continues the investigation of flux vacua on the Fermat sextic fourfold initiated in \cite{Braun:2020jrx}. While this fourfold does not carry an elliptic fibration and so cannot be used for F-Theory compactification, it's geometry is comparably simple, and enough is known about its middle 
cohomology to give a precise description of the questions we are interested in. It hence serves as an interesting arena to investigate the relationship
between the number of stabilized moduli and the associated tadpole cost. Fermat varieties and the associated Landau-Ginzburg models have 
been used for a detailed study of flux vacua in \cite{Bardzell:2022jfh,Becker:2022hse,Becker:2023rqi} based on the construction of \cite{Becker:2006ks}. 

We are focusing on supersymmetric Minkowski minima, where fluxes are characterised by primitive Hodge classes obeying \eqref{eq:fluxquant}, i.e. 
the set of fluxes is given by 
\begin{equation}
 \Lambda_{phys} = \left\{ G \in H^{2,2}(X)_{\prim}   \left| G + \frac{c_2(X)}{2} \in  H^4(X,\Z)\right. \right\} \, .
\end{equation}
Fluxes for which the rank of $\rho_{IJ}$ is maximal, so that all complex structure moduli are stabilized, are known as `general Hodge cycles' in the mathematics literature and we shall adopt this terminology here as well. Note that whenever 
$c_2(X)$ is odd, as for the sextic fourfold, $\Lambda_{phys}$ is not a lattice.

In the work \cite{Braun:2020jrx} general Hodge cycles constructed from linear algebraic cycles were studied. These form a rather special class of Hodge cycles, 
and we extend the scope of this work in two ways. We consider two different types of non-linear algebraic cycles in addition to linear algebraic cycles, and show how to use them to construct general Hodge cycles in Section \ref{sect:alg_cycles}.

Using algebraic cycles in our construction of fluxes tacitly assumes the integral Hodge conjecture. Although the set of algebraic cycles we consider are known to generate $H^{2,2}(X)\cap H^4(X,\Q)$, we show that they do not generate $ H^{2,2}(X) \cap H^4(X,\Z)$ in Section \ref{sect:hodge_conj}. We can however find a basis of $H^{2,2}(X) \cap H^4(X,\Z)$ using appropriately quantized Griffith residues, and show how to use this approach in Section \ref{sect:residues}. 
As it is computationally too expensive to write down general Hodge cycles using this approach, we restrict ourselves to models with specific symmetries. For the examples we consider, we show how to completely answer the question about the shortest general Hodge cycle. This setting also naturally leads to a consideration of quotients, and we comment on this possibility as well.  

Finally in Section \ref{sect:arithmetic} we showcase an alternative approach for investigating fluxes below the tadpole bound by using techniques from number theory to examine the quadratic form given by $\int_X G \wedge G$ on $ \Lambda_{phys}$.

To give some background and to fix notation we review those properties of the Fermat sextic that are needed in this work in Section \ref{sect:Fermat}. We have included tables of some of our results in an Appendix.

\section{The Fermat Sextic Fourfold}\label{sect:Fermat}
The Fermat sextic fourfold $X$ is the hypersurface
\begin{equation}
Q(x) \equiv x_0^6 + x_1^6 + x_2^6 + x_3^6 + x_4^6 + x_5^6 = 0 
\end{equation}
in $\P^5$ with homogeneous coordinates $[x_0:x_1:x_2:x_3:x_4:x_5]$. It is well known that its group of automorphisms is 
$$
\Aut(X)=\mathfrak{S}_{6}\ltimes \Z_6^5
$$
where $\mathfrak{S}_6$ is the group of permutations of six elements, which acts on $X$ by permutation of coordinates, and we identify
$$
\Z_6^5\simeq \Z_6^6/D
$$
where $D$ is the diagonal
$$
D\equiv \text{Im}(a\in \Z_6\mapsto (a,a,a,a,a,a)\in\Z_6^6).
$$
The group $\Z_6^6/D$ acts on $X$ as
$$
\ell:[x_0:x_1:x_2:x_3:x_4:x_5]\mapsto [\zeta^{\ell_0}x_0:\zeta^{\ell_1}x_1:\zeta^{\ell_2}x_2:\zeta^{\ell_3}x_3:\zeta^{\ell_4}x_4:\zeta^{\ell_5}x_5],
$$
with $\zeta=e^{\frac{\pi i}{3}}$ the primitive sixth root of unity.
%\ab{state some more basic properties} 

The total Chern class of $X$ is by adjunction
\begin{equation}
\label{eq:totalchern}
    c(X) = (1+H)^6/(1+6H) = 1 + 15 H^2 - 70 H^3 + 435 H^4 \, ,
\end{equation}
where $H\in H^{1,1}(X)\cap H^2(X,\Z)$ is the hyperplane class. This
implies that $\chi(X) = 2610$ and so the tadpole bound corresponds to
\begin{equation}\label{eq:tadpolesextic}
\int_X G\wedge G\le 217.5.
\end{equation}
The second Chern class of $X$ is odd and so the quantization condition for 
$\Lambda_{phys}$ is 
\begin{equation}\label{eq:quantcondsextic}
G + \frac{1}{2} H^2 \in H^4(X,\Z)\, .
\end{equation}

\subsection{Cohomology of the Fermat sextic}

For any two elements of $H^4(X)$ we abbreviate 
\begin{equation}
A \cdot B := \int_X A \wedge B \, . 
\end{equation}
The primitive part of the middle cohomology group of $X$ corresponds to
$$
H^4(X)_\prim= \{A\in H^4(X) | A\cdot H^2=0\} \, ,
$$
and can be described by means of the residue map
$$
Res: H^5(\P^5\setminus X)\rightarrow H^4(X).
$$
The residue mapping is surjective onto the primitive middle cohomology and, after the work of Griffiths \cite{gr69}, a basis of $H^4(X)_\prim$ is given by the forms
\begin{align*}
\omega_{\beta} := Res\left(\frac{x^{\beta}\Omega_0}{Q(x)^{k+1}}\right)
\end{align*}
where 
$$
\Omega_0=\sum_{i=0}^5(-1)^ix_i dx_0\wedge\cdots\widehat{dx_i}\cdots\wedge dx_5
$$
is the standard degree 6 top form of $\P^5$, $x^{\beta}$ is the monomial
\begin{align*}
    x^{\beta}=x_0^{\beta_0}x_1^{\beta_1}x_2^{\beta_2}x_3^{\beta_3}x_4^{\beta_4}x_5^{\beta_5}
\end{align*}
with $|\beta| := \tfrac{1}{6}\sum \beta_i = k \in \mathbb{Z}$, $0\leq \beta_i \leq 4$, and $0 \leq k \leq 4$ determines the Hodge type: 
\begin{align*}
\omega_{\beta} \in H^{4-k,k}(X)\, . 
\end{align*}

The inner form between any two residues is found from the following statement taken from \cite[Proposition 8.4]{ML}:
For any two monomials $P(x)$ and $R(x)$ of degrees $6p$ and $6q$ such that $p+q=4$ we have 
\begin{equation}\label{eq:intgriffbasis}
\omega_P \cdot\omega_R  =   \int_X \omega_P \wedge \omega_R = c\,\, (-1)^{p+1} \,\, \frac{(2\pi i)^4}{p!q!}5^6 6
\end{equation}
for $c\in\C$ the unique number such that 
\begin{equation}
PR \equiv c \det(\mbox{Hess}(Q))\,\, \, \, \, \, (\mbox{mod} \,\, \mbox{Jac}(Q)) ,
\end{equation} 
where $\text{Jac}(Q):=\langle \frac{\partial Q}{\partial x_0},\ldots,\frac{\partial Q}{\partial x_5}\rangle$ is the Jacobian ideal of $Q(x)$, and $\text{Hess}(Q)$ is its Hessian matrix.

For the Fermat sextic the Hessian determinant is simply
\begin{equation}
\det(\mbox{Hess}(Q)) = 30^6 \prod_i x_i^4 \, ,
\end{equation}
and so for our monomial basis
\begin{equation}
 \omega_{\beta} \cdot \omega_{\beta'} = 0
\end{equation}
except when $\beta_i = 4 - \beta_i'$ for all $i=0,1,2,3,4,5$. We hence define 
\begin{equation}
\bar{\beta}_i := 4-\beta_i 
\end{equation}
which implies that
\begin{equation}
|\bar{\beta}| = \tfrac16 \sum_{i=0}^5 (4-\beta_i) = 4- |\beta| .
\end{equation}

Since $x^{\beta} x^{\bar{\beta}}= \prod_i x_i^4$, it follows that $c= 30^{-6}$ and 
\begin{equation}\label{eq:inner_form_residues}
\omega_{\beta} \cdot \omega_{\bar{\beta}} =  (-1)^{|\beta|+1} \left(\frac{1}{30}\right)^6 \frac{(2\pi i)^4}{|\beta|!|\bar{\beta}|!} 5^6 \cdot 6 
=  (-1)^{|\beta|+1}  \frac{(2\pi i)^4}{|\beta|!|\bar{\beta}|!} \frac{1}{ 6^5}.
\end{equation}

\subsection{Hodge and Algebraic cycles}\label{subsect:Fermat-Hodgealgcyc}

We will call classes in $H^{2,2}(X) \cap H^4(X,\Q)$ Hodge cycles, and classes in 
$H^{2,2}(X) \cap H^4(X,\Z)$ integral Hodge cycles. After the work of Shioda \cite{sh79} we know that the residue forms generate the space of complexified primitive Hodge cycles
$$
H^{2,2}(X)_\prim=(H^{2,2}(X)_\prim\cap H^4(X,\Z))\otimes\C, 
$$
this can also be expressed as $X$ has maximal Hodge rank.

For $|\beta|=2$, we have that $\omega_{\beta} \in  H^{2,2}(X)_\prim$. Such a form $\omega_{\beta}$ will be called: 
\begin{itemize}
    \item 3-decomposable if $\beta=(a,4-a,b,4-b,c,4-c)$ 
    \item 1-decomposable if $\beta=(a,4-a,0,2,3,3)$
    \item indecomposable if $\beta=(0,0,3,3,3,3)$
\end{itemize}
up to permutations and for $0 \leq a, b, c \leq 4$. The $1751$ classes in $H^{2,2}(X)_{\prim}$ are thus organized into $1001$ 3-decomposable, $720$ 1-decomposable and $30$ indecomposable cycles.

On the other hand it was shown by Shioda \cite{sh79} that $X$ satisfies the Hodge conjecture over $\Q$. In fact, we will provide 1751 linearly independent primitive algebraic cycles inside $X$ which can be divided into 3 different types corresponding to the three types of residues defined above. 

\begin{itemize}
    \item Linear cycles: which can be obtained as the orbit of $\Aut(X)=\mathfrak{S}_6\ltimes \Z_6^5$ on
    \begin{equation}
    \label{eq:lincyc0}
    C:= \{x_0-\mu x_1=x_2-\mu x_3=x_4-\mu x_5=0\}\subseteq X,
    \end{equation}
    where $\mu=e^{\frac{\pi i}{6}}$ is the primitive 12th root of unity (and so $\mu^2=\zeta$). Given $\sigma\in \mathfrak{S}_6$ and $\ell\in \Z_6^6/D\simeq \Z_6^5$ we denote by
    $$
    C^\ell_\sigma:= \sigma^{-1}(\ell^{-1}(C)) \, ,
    $$
    which is explicitly given by the equations 
    \begin{equation}
    \label{eq:lincyc}
    x_{\sigma(0)}-\mu^{2(\ell_1-\ell_0)+1}x_{\sigma(1)}=x_{\sigma(2)}-\mu^{2(\ell_3-\ell_2)+1}x_{\sigma(3)}=x_{\sigma(4)}-\mu^{2(\ell_5-\ell_4)+1}x_{\sigma(5)}=0.
    \end{equation}
    \item Aoki-Shioda cycles: which are obtained as the orbit of $\Aut(X)$ on
    \begin{equation}
    \label{eq:AScyc0}
    S:= \{x_0^2-\sqrt[3]{2}x_1x_2=x_1^3+x_2^3+ix_3^3=x_4-\mu x_5=0\}\subseteq X.
    \end{equation}
    Similarly we denote $S^\ell_\sigma:= \sigma^{-1}(\ell^{-1}(S))$ for $\sigma\in \mathfrak{S}_6$ and $\ell\in\Z_6^6/D$. It is given by the equations
$$
\zeta^{2\ell_0}x_{\sigma(0)}^2-\sqrt[3]{2}\zeta^{\ell_1+\ell_2}x_{\sigma(1)}x_{\sigma(2)}=0,
$$
\begin{equation}
\label{eq:AScyc}
(-1)^{\ell_1}x_{\sigma(1)}^3+(-1)^{\ell_2}x_{\sigma(2)}^3+i(-1)^{\ell_3}x_{\sigma(3)}^3=0,
\end{equation}
$$
\zeta^{\ell_4}x_{\sigma(4)}-\mu^{1+2\ell_5}x_{\sigma(5)}=0.
$$

    \item Type 3 cycles: which are in the orbit of $\Aut(X)$ on
    \begin{equation}
    \label{eq:T3cyc0}
    T:= \{x_0^2-\sqrt[3]{2}x_1x_2=x_3^2-\sqrt[3]{2}x_4x_5=x_1^3+x_2^3+ix_4^3+ix_5^3=0\}\subseteq X.
    \end{equation}
    We denote $T^\ell_\sigma:= \sigma^{-1}(\ell^{-1}(T))$, where $\sigma\in\mathfrak{S}_6$ and $\ell\in\Z_6^6/D$. The explicit equations of $T_\sigma^\ell$ are
    $$
     \zeta^{2\ell_0}x_{\sigma(0)}^2 -\sqrt[3]{2}\zeta^{\ell_1+\ell_2}x_{\sigma(1)}x_{\sigma(2)}=0, 
     $$
     \begin{equation}
     \label{eq:T3cyc}    
     \zeta^{2\ell_3}x_{\sigma(3)}^2 -\sqrt[3]{2}\zeta^{\ell_4+\ell_5}x_{\sigma(4)}x_{\sigma(5)}=0,
     \end{equation}
     $$ (-1)^{\ell_1}x_{\sigma(1)}^3+(-1)^{\ell_2}x_{\sigma(2)}^3+i(-1)^{\ell_4}x_{\sigma(4)}^3+i(-1)^{\ell_5}x_{\sigma(5)}^3=0.
     $$
\end{itemize}

After \cite[Theorem 1.2]{villaflor2021periods} it is possible to compute all periods of residue forms over such algebraic cycles. For the particular case of linear cycles, an explicit formula was obtained in \cite{movasati_loyola_17} and is the following:

\begin{equation}\label{eq:movisato_loyola_p2intersections}
\int_{C_\sigma^{\,\boldsymbol{\ell}}} \omega_\beta=
\left\{
\begin{array}{ll}
(2 \pi i)^2 \frac{{\rm sgn}(\sigma)}{6^3 \cdot 2} \,\mu^{  \sum_{e=0}^2 (\beta_{\sigma(2e)}+1)(2(\ell_{2e+1}-\ell_{2e})+1) } & \mbox{if }\, 
\beta_{\sigma(2e-2)}+\beta_{\sigma(2e-1)} = 4 \\ \\
0 & \mbox{otherwise.}\\
\end{array}\right.
\end{equation}

\subsection{Hodge loci}\label{subsect:Hodgeloci}

Given a Hodge cycle $G$, the Hodge locus of $G$ is the germ of the analytic subvariety of the parameter space of sextic fourfolds $\T=H^0(\mathcal{O}_{\P^5}(6))$ given by
\begin{equation}
V_G:= \{t\in (\T,t_0)| G_t\in H^{2,2}(X_t)\cap H^4(X_t,\Z)\} \, ,
\end{equation}
where $X_t$ is the hypersurface defined by the sextic $t\in \T$, $t_0=Q(x)$ is the defining equation of the Fermat sextic, $G_{t_0}=G$ and $G_t\in H^4(X_t,\Z)$ is the class obtained by monodromy in a small analytic neighbourhood $t\in (\T,t_0)$.

Using the infinitesimal variation of Hodge structure
$$
\nabla: T_{t_0}\T \rightarrow \hom(H^{3,1}(X),H^{2,2}(X))
$$
composed with the multiplication by $G$ map
$$
\varphi_G: H^{2,2}(X)\rightarrow \C
$$
$$
\eta\mapsto G\cdot \eta
$$
we can describe the Zariski tangent space of $V_G$ as
\begin{equation}
T_{t_0}V_G=\ker(\varphi_G^*\circ \nabla: T_{t_0}\T\rightarrow (H^{3,1}(X))^\vee).    
\end{equation}
Since $U$ is a linear space, every tangent vector $v\in T_{t_0}\T=\T$ is a degree 6 polynomial in the variables $x=(x_0,\ldots,x_5)$. Under this identification we can express
$$
(\varphi_G^*\circ \nabla)(v): H^{3,1}(X)\rightarrow \C
$$
$$
\omega_{\beta}\mapsto Res\left(\frac{v(x)x^{\beta}\Omega_0}{Q(x)^3}\right)\cdot G
$$
where $|\beta|=1$.

For a Hodge cycle $G$ we define 
 \begin{equation}\label{eq:rhomatrix}
\rho_{IJ}(G) :=  \omega_{\beta_I+\beta_J} \cdot  G 
 \end{equation}
for $|\beta_I|=|\beta_J|=1$. Denoting the associated $426\times426$ square matrix by $\rho(G):= (\{\rho_{IJ}(G)\}_{I,J})$, it follows that
\begin{equation}
\text{rank } \rho(G)=\text{Codim}(T_{t_0}V_G\subseteq T_{t_0}\T) \, .
\end{equation}
We say that $G$ is a general Hodge cycle if 
\begin{equation}\label{eq:genHodge}
\text{Codim}(V_G\subseteq \T)=h^{3,1}(X)=426.    
\end{equation}
In particular this happens if $\rho(G)$ is of full rank (i.e. if $\rho(G)$ is invertible) and in such a case the corresponding Hodge locus $V_G$ is smooth and reduced as an analytic scheme. We remark that this is not a characterization of general Hodge cycles, since there might be some singular or non-reduced Hodge locus $V_G$ for some general Hodge cycle, in those cases $\rho(G)$ fails to be invertible. 

In general, given a Hodge cycle $G\in H^{2,2}(X)\cap H^4(X,\Q)$ we can use the information given by the periods of residue forms over it to study its Hodge locus. 

\subsection{Vanishing cycles}

Let us turn now to the middle homology group of the Fermat sextic. The affine Fermat sextic is given by 
$$
U_0:=\{(x_1,x_2,x_3,x_4,x_{5})\in \C^{5} | \ 1+x_1^6+x_2^6+x_3^6+x_4^6+x_{5}^6=0\}\, .
$$
A basis for $H_4(U_0,\Z)$ is given by the so called vanishing cycles, which can be defined as follows: For every $\beta\in\{0,1,2,3,4\}^{5}$ consider the homological cycle
$$
\delta_{\beta}:=\sum_{a\in\{0,1\}^{5}}(-1)^{\sum_{i=1}^{5}(1-a_i)}\Delta_{\beta+a}
$$
where $\Delta_{\beta+a}:\Delta^4:=\{(t_1,t_2,t_3,t_4,t_{5})\in\R^{5}: t_i\ge 0 \ , \ \sum_{i=1}^{5}t_i=1\}\rightarrow U_0$ is given by
$$
\Delta_{\beta+a}(t):=\left(\zeta_{12}^{2(\beta_1+a_1)-1}t_1^\frac{1}{6},\zeta_{12}^{2(\beta_2+a_2)-1}t_2^\frac{1}{6},\zeta_{12}^{2(\beta_3+a_3)-1}t_3^\frac{1}{6},\zeta_{12}^{2(\beta_4+a_4)-1}t_4^\frac{1}{6},\zeta_{12}^{2(\beta_{5}+a_{5})-1}t_{5}^\frac{1}{6}\right).
$$
The set $\{\delta_\beta\}_{\beta\in\{0,1,2,3,4\}^{5}}$ is a basis of $H_4(U_0,\Z)$.

Using the Leray-Thom-Gysin sequence in homology \cite[\S 4.6]{ho13} is easy to see that
\begin{equation}
\label{lefsdescrathom}    
H_4(X,\Q)=H_4(X,\Q)_\prim\oplus\Q\cdot H^2
\end{equation}
where 
$$
H_4(X,\Q)_\prim:=\text{Im}(H_4(U_0,\Q)\rightarrow H_4(X,\Q))\, ,
$$
and $H^2$ is the class of the square of the hyperplane section of $\P^5$ intersected with $X$. Hence every $\omega\in H^4(X)$ is determined by its periods over the vanishing cycles and $H^2$. Since this last period is zero when $\omega\in H^4(X)_\prim$, we see that every primitive class is determined by its periods over all vanishing cycles. These periods can be explicitly computed \cite{dmos} as follows: 
\begin{align}\label{eq:period_vancyc_residue}
\int_{\delta_{\beta'}}\omega_{\beta}=\frac{{(-1)^{|\beta|}}}{6^5} \frac{1}{|\beta|! 2 \pi i } \prod_{i=0}^5 \Gamma\left(\frac{\beta_i+1}{6} \right) 
    \left(\zeta^{(\beta'_i+1)(\beta_i+1)}-\zeta^{(\beta'_i)(\beta_i+1)}\right)
\end{align}
where $\beta_0':=0$.

Similarly one can produce another basis of vanishing cycles for the affine Fermat sextic 
$$
U_i:=X\cap\{x_i=1\}
$$ 
and get the same formula where $\beta_i':=0$. Hence in general a vanishing cycle is represented by a tuple 
\begin{align*}
    \delta_{\beta}=[\beta_0,\beta_1,\beta_2,\beta_3,\beta_4,\beta_5]
\end{align*}
where each of the $\beta_i$ is an integer ranging from $0$ to $4$, one of the $\beta_i$ must be $0$, and its pairing with a primitive cohomology class is given by \eqref{eq:period_vancyc_residue}.

The decomposition \eqref{lefsdescrathom} is not true anymore over $\Z$. In fact, looking at the Leray-Thom-Gysin sequence in homology \cite[\S 4.6]{ho13} we get the short exact sequence
$$
0\rightarrow H_4(X,\Z)_\prim\rightarrow H_4(X,\Z)\xrightarrow{f} H_{2}(X_\infty,\Z)\rightarrow 0
$$
where $H_4(X,\Z)_\prim:=\text{Im}(H_4(U_0,\Z)\rightarrow H_4(X,\Z))$, $f$ is the intersection map and
$$
X_\infty=X\cap\{x_0=0\}=\{x_1^6+x_2^6+x_3^6+x_4^6+x_5^6=0\}
$$ 
is the Fermat sextic threefold at infinity. Since 
$$
H_{2}(X_\infty,\Z)=\Z\cdot[L]
$$
for some line $L\subseteq X_\infty$, we get the following decomposition
\begin{equation}
\label{lefsdecinthom}
H_4(X,\Z)=H_4(X,\Z)_\prim\oplus \Z\cdot[C]    
\end{equation}
for any linear algebraic cycle $C\subseteq X$. We remark that the decomposition \eqref{lefsdecinthom} is not orthogonal with respect to the intersection pairing, while \eqref{lefsdescrathom} is. Thus it is not true in general that the primitive part (i.e. its orthogonal projection to the primitive subspace) of an integral cycle is integral, but this will be the case if it has degree divisible by $6$, as for the Aoki-Shioda and type 3 cycles.

We remark that since there are no torsion cycles in the Fermat sextic, we have duality between integral homology and cohomology
$$
H^4(X,\Z)=(H_4(X,\Z))^\vee.
$$

\section{The Hodge Conjecture}\label{sect:hodge_conj}
Before treating the tadpole conjecture, in this section we will discuss an explicit approach to the Hodge conjecture for the Fermat sextic fourfold. Our objective is to exhibit a basis of algebraic cycles which generate the space of Hodge cycles and then use it to look for possible representatives of the G-flux in the next section. In spite that the Hodge conjecture over $\Q$ for the Fermat sextic holds by the work of Shioda \cite{sh79}, the integral Hodge conjecture is still open and so it might happen that the G-flux is an algebraic cycle divided by a natural number. 

As explained in Section \ref{subsect:Fermat-Hodgealgcyc}, the space of primitive algebraic cycles has dimension 1751. Moreover, any primitive Hodge cycle
$$
\omega_P=Res\left(\frac{P(x)\Omega_0}{Q(x)^3}\right)\in H^{2,2}(X)_\prim\cap H^4(X,\Q)
$$
is linear combination of residue forms $\omega_\beta$ which can be 3-decomposable (generating a space of dimension 1001), 1-decomposable (generating a space of dimension 720) or indecomposable (generating a space of dimension 30). Using the main theorem of \cite[Theorem 1.1]{villaflor2021periods} we can write this linear combination for the primitive part of all the algebraic cycles described in Section \ref{subsect:Fermat-Hodgealgcyc} as follows: 

\begin{itemize}
    \item For the linear cycle $C$ given by \eqref{eq:lincyc0} we have $[C]_\prim=\omega_P$ for 
    \begin{equation}
    P=-72i\left(\sum_{j=0}^4x_0^{j}(\mu x_1)^{4-j}\right)\left(\sum_{j=0}^4x_2^{j}(\mu x_3)^{4-j}\right)\left(\sum_{j=0}^4x_4^{j}(\mu x_5)^{4-j}\right).
    \end{equation}
    This polynomial is a linear combination of 3-decomposable residue forms.
    \item For the Aoki-Shioda cycle $S$ given by \eqref{eq:AScyc0} we have $[S]_\prim=\omega_P$ for
    \begin{equation}
    P=-6^3\cdot \sqrt[3]{2}\mu^4 x_0x_3^2(x_0^2+\sqrt[3]{2}x_1x_2)(x_1^3-x_2^3)\left(\sum_{j=0}^4x_4^j(\mu x_5)^{4-j}\right),
    \end{equation}
    which is a linear combination of 1-decomposable residue forms.
    \item For the type 3 cycle $T$ given by \eqref{eq:T3cyc0} we have $[T]_\prim=\omega_P$ for
    \begin{equation}
    P=-6^3 \cdot\sqrt[3]{4}ix_0x_3(x_0^2+\sqrt[3]{2}x_1x_2)(x_3^2+\sqrt[3]{2}x_4x_5)(x_1^3-x_2^3)(x_5^3-x_4^3),
    \end{equation}
    which is a combination of 3-decomposable and indecomposable residue forms.
\end{itemize}
Since the action of $\Aut(X)=\mathfrak{S}_6\ltimes \Z_6^5$ on the space of residue forms preserves the spaces of 3-decomposable, 1-decomposable and indecomposable forms respectively, the primitive part of all linear cycles $C^\ell_\sigma$ in the orbit of $C$ are explicit linear combinations of 3-decomposable residue forms. Similarly Aoki-Shioda cycles $S^\ell_\sigma$ are linear combinations of 1-decomposable forms, while type 3 cycles $T^\ell_\sigma$  are linear combinations of 3-decomposable and indecomposable forms. 

Using these explicit combinations it is possible to compute (using \eqref{eq:intgriffbasis}) the intersection matrix of all linear cycles, and extract a subset of maximal rank 1001. Therefore, these linear cycles generate the space of all 3-decomposable residue forms. Similarly for Aoki-Shioda cycles we can find 720 of them whose intersection matrix is of full rank, and so they generate all 1-decomposable residue forms. For type 3 cycles, since they also have a 3-decomposable part, we have to look for 30 cycles whose intersection matrix together with the 1001 linear cycles has full rank 1031. We implemented these algorithms in the {\sc Singular} library \texttt{"HodgeProject.lib"}, as the procedures \texttt{Periods}, \texttt{IntersecMatrix}, \texttt{LinAlgCycles}, \texttt{ASAlgCycles} and \texttt{T3AlgCycles} \footnote{https://github.com/danfelmath01/Sixtic-Fourfold}. We summarize the obtained basis in the following statement.

%\ab{should put a link/reference for the libraries used}

\begin{theorem}
\label{thmHC}
The Fermat sextic fourfold $X$ satisfies the Hodge conjecture. In fact, a basis for $H^{2,2}(X)\cap H^4(X,\Q)_\prim$ is given by the primitive part of 1001 linear cycles, 720 Aoki-Shioda cycles and 30 type 3 cycles. This set of algebraic cycles can be computed explicitly and we denote it $\texttt{Alg}=\{A_1,\ldots,A_{1751}\}=\{C_1,\ldots,C_{1001},S_1,\ldots,S_{720}, T_1,\ldots, T_{30}\}$. 
\end{theorem}

In what follows we will list the 1751 algebraic cycles according to the action of the corresponding pair $(\sigma,\ell)\in \mathfrak{S}_6\ltimes\Z_6^5=\Aut(X)$. 

\subsection{Linear cycles}

In view of the explicit equation of the linear cycle $C_\sigma^\ell$ given by \eqref{eq:lincyc}, we can avoid some repeated terms if we set $\ell_0=\ell_2=\ell_4=0$. After obtaining the part of the set $\texttt{Alg}$ given by the first 1001 linear cycles, it turns out that the permutations of $\{0,1,2,3,4,5\}$ associated with them are all contained in the following list of 15 permutations: 
\begin{equation}\label{eq:sigma}
\begin{aligned}
\Sigma=&(0,1,2,3,4,5),(0,2,1,3,4,5),(0,3,1,2,4,5),(0,1,2,4,3,5), (0,2,1,4,3,5),\\
&(0,4,1,2,3,5),(0,3,1,4,2,5),(0,4,1,3,2,5),(0,1,2,5,3,4),(0,2,1,5,3,4),\\
&(0,5,1,2,3,4),(0,3,1,5,2,4),(0,5,1,3,2,4),(0,4,1,5,2,3), (0,5,1,4,2,3).
\end{aligned}
\end{equation}
The full list of linearly independent linear cycles $C_i=A_i$ for $i=1,\ldots,1001$ is presented in Table \ref{table:linear cycles} in Appendix \ref{app:lin_indep_alg_cycles}.

\subsection{Aoki-Shioda cycles}

For Aoki-Shioda cycles, the explicit equation of $S_\sigma^\ell$ is \eqref{eq:AScyc}. In order to avoid repetitions we set $\ell_1=\ell_4=0$, and restrict the set of permutations of $\{0,1,2,3,4,5\}$ to the set $\Sigma'$ consisting of those satisfying $\sigma(1)<\sigma(2)$. The list of linearly independent Aoki-Shioda cycles $S_i=A_{i+1001}$ for $i=1,\ldots,720$ is presented in Table \ref{table:AS cycles} in Appendix \ref{app:lin_indep_alg_cycles}.

\subsection{Type 3 cycles}

The explicit equation of type 3 cycles $T_\sigma^\ell$ is given by \eqref{eq:AScyc}. In order to avoid repetitions we set $\ell_0=0$, and restrict the set of permutations of $\{0,1,2,3,4,5\}$ to the set $\Sigma''$ consisting of those satisfying that $\sigma(1)<\sigma(2)$ and $\sigma(4)<\sigma(5)$. The list of linearly independent type 3 cycles $T_i=A_{i+1721}$ for $i=1,\ldots,30$ is presented in Table \ref{table:t3 cycles} in Appendix \ref{app:lin_indep_alg_cycles}.

\subsection{The Integral Hodge Conjecture}

Given our basis of algebraic cycles $\texttt{Alg}$, one can wonder how far are we from the lattice of primitive integral Hodge cycles. Using the period equations \eqref{eq:period_vancyc_residue}, it is possible implement and find a dual basis for $H^{2,2}(X)_\prim\cap H^4(X,\Z)$ in terms of vanishing cycles. On the other hand, using \eqref{eq:intgriffbasis} it is possible to compute the period matrix of $\texttt{Alg}$ and compare it to the period matrix of the basis of primitive integral Hodge cycles obtaining the following result.

\begin{theorem}
The Fermat sextic fourfold $X$ satisfies the $\Z[\frac{1}{6}]$-Hodge conjecture, this means
$$
H^{2,2}(X)\cap H^4(X,\Z[6^{-1}])=H^4(X,\Z[6^{-1}])_{\text{alg}}.
$$
More precisely, if $\texttt{Alg}=\{A_1,\ldots,A_{1721}\}$ is the basis of algebraic cycles of Theorem \ref{thmHC}, then
$$
H^{2,2}(X)\cap H^4(X,\Z)_\prim\subseteq\bigoplus_{i=1}^{1721}\Z\cdot\frac{1}{d_i}[A_i]_\prim
$$
where each $d_i\in\{1,2,3,6\}$. We have listed the $d_i$ appearing in table \ref{table:divisor of alg cycles}.

\begin{table}[!ht]
\tiny
\centering
\begin{tabular}{|l|l|}
		\hline
		Index& Divisors \\
		\hline
		\hline
		1, \ldots, 60 &6,3,6,6,3,6,3,6,3,3,  6,6,6,3,6,3,3,3,6,3,  6,3,6,6,3,6,3,3,3,3,  3,3,3,6,3,6,6,3,6,3,  6,6,6,6,6,6,3,3,3,3,  3,6,6,6,3,6,6,3,6,3\\
        61, \ldots, 120& 6,3,6,3,3,6,3,6,6,3, 6,6,6,3,6,6,6,3,3,6, 3,3,6,6,3,6,6,3,3,6, 6,6,6,6,6,6,3,3,3,3, 3,6,3,3,6,3,3,6,3,3, 3,6,3,3,3,6,6,3,6,3\\
        121,\ldots, 180& 3,3,6,6,3,3,3,6,6,3, 3,3,6,6,6,6,6,6,6,6, 3,3,3,3,3,6,6,3,3,3, 6,6,3,6,6,6,3,3,3,3, 3,6,3,3,6,3,6,6,3,6, 3,6,3,3,3,6,6,6,3,6\\
        181,\ldots, 240 & 3,3,6,3,3,3,3,3,3,3, 6,6,6,6,3,3,3,6,6,3, 6,6,6,6,3,6,3,3,6,3, 3,6,6,6,6,6,6,6,6,3, 6,6,3,3,6,3,6,3,6,6, 6,3,6,6,3,3,3,3,6,3\\
        241,\ldots, 300 &  3,6,3,3,6,3,3,3,3,3, 3,6,3,6,3,3,6,6,6,6, 3,3,6,6,3,3,3,3,3,6, 3,6,6,3,6,6,3,3,3,3, 6,6,3,6,3,3,6,3,3,6, 6,6,3,3,3,3,3,6,6,1\\
        301,\ldots, 360& 1,6,1,6,3,3,6,6,6,3, 6,6,6,3,6,3,6,3,3,6, 6,6,6,6,3,6,6,3,3,6, 6,6,3,3,6,6,3,6,3,3, 6,3,3,6,6,6,6,3,3,3, 3,3,3,3,3,3,3,3,6,6\\
        361, \ldots, 420 & 6,6,6,3,6,3,6,6,6,3, 6,6,3,3,3,6,3,6,3,6, 3,3,6,3,3,6,3,3,6,6, 6,6,3,6,3,3,6,3,6,6, 3,3,3,3,3,3,3,6,6,6, 6,6,6,6,6,3,3,3,6,3\\
        421,\ldots, 480&        6,3,3,6,6,3,6,6,6,6, 3,3,3,6,6,6,6,6,3,6, 3,6,6,6,3,6,6,3,6,6, 3,6,3,3,3,6,6,3,6,3, 6,6,3,3,3,6,3,6,3,6, 3,6,6,3,6,6,6,3,3,6\\
        481,\ldots, 540 & 6,6,3,6,6,3,6,6,6,6, 3,3,3,3,3,3,3,6,6,3, 3,3,3,3,6,6,6,3,3,6, 6,6,3,3,3,3,3,3,6,6, 3,1,1,3,3,3,3,3,6,6, 3,6,6,6,6,3,3,3,3,3\\
        541,\ldots, 600 & 3,3,6,3,6,6,3,6,3,3, 3,3,3,6,6,6,6,3,6,6, 3,6,6,3,3,3,3,3,3,6, 3,6,6,3,6,6,3,6,3,3, 3,3,3,3,3,3,3,3,6,3, 3,6,3,6,6,3,3,3,6,6\\
        601,\ldots, 660& 6,6,3,6,3,3,6,3,3,3, 3,3,3,6,6,6,6,1,1,3, 1,3,3,3,3,3,3,6,6,3, 3,3,3,3,3,6,3,6,6,3, 6,6,6,3,3,3,6,3,6,3, 6,3,3,3,3,3,3,3,6,3\\
        661,\ldots, 720& 3,3,6,3,6,6,3,6,6,3, 6,6,6,3,3,3,6,3,6,3, 6,6,6,3,3,6,6,3,6,6, 3,3,6,6,6,3,6,6,3,3, 6,6,3,3,6,3,6,3,3,3, 3,3,3,6,6,3,3,6,6,6\\
        721,\ldots, 780& 3,3,3,6,3,6,6,3,6,6, 3,3,3,3,3,3,6,3,6,3, 6,3,6,3,6,6,6,3,6,6, 3,3,6,6,6,3,6,3,3,6, 3,3,3,3,3,3,3,6,3,6, 3,6,3,3,6,3,3,3,3,3\\
        781,\ldots, 840& 3,3,3,3,3,3,3,6,3,3, 6,6,6,1,3,3,6,6,6,6, 3,6,6,3,6,6,3,3,3,3, 6,6,6,6,3,6,3,3,6,3, 6,6,3,3,3,6,3,6,6,6, 3,6,6,3,3,3,3,3,6,3\\
        841,\ldots, 900& 6,3,6,3,6,6,3,6,6,6, 3,3,6,6,6,6,6,6,3,6, 6,3,6,6,3,3,3,3,3,3, 3,6,3,6,6,3,6,6,6,3, 3,3,6,6,6,6,3,6,3,3, 6,3,3,3,3,3,3,3,6,3\\
        901,\ldots, 960& 6,3,3,3,3,3,3,3,3,3, 6,3,6,3,3,6,3,6,6,6, 3,3,3,6,3,6,3,1,1,3, 1,6,6,3,3,1,3,1,3,3, 1,6,3,6,1,6,1,6,6,6, 6,6,3,3,1,3,6,6,3,6\\
        961,\ldots, 1001& 6,6,1,6,3,1,3,3,1,1, 6,3,6,1,1,1,1,6,6,6, 6,3,1,1,3,6,3,6,3,6, 3,6,3,3,1,1,1,1,1,6, 6\\
        \hline
      1002,\ldots, 10061& 3,6,6,3,3,3,3,6,6,6, 6,6,3,6,6,3,3,3,3,6, 6,6,6,6,3,6,6,3,6,6, 3,6,2,6,6,2,3,6,6,3, 6,6,3,3,3,6,6,2,3,3, 3,3,3,3,3,3,3,1,1,3\\
       1062,\ldots, 1121& 3,3,3,3,3,3,3,3,3,3, 3,3,6,6,6,3,3,3,6,3, 3,6,6,3,6,6,6,6,3,3, 6,3,1,6,6,1,3,6,6,6, 3,3,3,3,3,6,6,1,2,6, 6,3,3,3,3,3,3,1,1,3\\
       1122,\ldots, 1181& 3,3,3,3,3,3,3,3,3,3, 3,3,6,6,6,3,3,3,6,3, 3,6,6,3,6,6,6,6,3,3, 2,1,1,2,1,1,6,2,2,6, 1,1,1,1,1,2,1,1,6,2, 2,3,1,1,1,1,1,1,1,1\\
       1182,\ldots, 1241& 3,3,3,3,3,3,3,3,3,3, 3,3,6,6,6,6,1,3,6,3, 3,6,6,3,6,6,6,6,3,3, 2,1,2,1,6,2,2,6,1,1, 2,1,1,2,1,1,6,2,2,3, 1,1,1,1,1,1,1,1,3,3\\
       1242,\ldots, 1301& 3,3,3,3,3,3,3,1,3,3, 3,6,6,6,3,3,3,3,3,2, 6,3,6,6,6,6,3,3,1,1, 2,1,6,1,6,1,2,1,2,6, 2,2,3,1,1,1,1,1,1,1, 1,3,3,3,3,3,3,3,3,3\\
       1302,\ldots, 1361& 1,1,3,1,3,3,3,3,3,3, 3,3,1,1,3,3,1,1,3,1, 1,1,1,1,1,3,1,3,1,1, 1,1,3,3,1,1,3,6,6,3, 3,3,3,6,6,6,6,6,3,6, 6,3,3,3,3,6,6,6,6,6\\
       1362,\ldots, 1421& 3,6,6,3,6,6,3,6,6,6, 6,6,3,6,6,3,6,6,3,3, 3,6,2,6,3,3,3,3,3,3, 3,3,3,1,1,3,3,6,6,3, 3,3,3,6,6,6,6,6,6,6, 6,3,3,3,3,3,3,3,3,3\\
       1422,\ldots, 1481& 6,6,6,3,3,3,3,3,3,3, 3,3,6,6,3,3,3,3,3,3, 3,3,1,3,1,3,3,3,3,3, 3,3,3,1,3,3,6,6,3,3, 3,3,6,6,6,6,6,6,6,6, 3,3,3,3,3,3,6,6,6,3\\
       1482,\ldots, 1541& 3,3,1,1,1,1,1,1,6,6, 3,3,3,3,1,1,1,1,1,3, 1,1,3,1,1,1,1,3,6,6, 3,6,6,3,2,2,6,2,2,6, 6,6,3,3,1,3,1,3,6,6, 6,3,1,6,2,3,1,1,1,1\\
       1542,\ldots, 1601& 1,1,3,1,3,1,1,3,6,6, 1,6,2,1,1,6,2,2,2,6, 1,3,1,1,3,2,2,1,2,2, 3,3,1,1,3,1,1,1,1,1, 1,1,1,1,1,1,1,1,1,1, 1,1,3,3,3,3,3,3,3,3\\
       1602,\ldots, 1661& 3,3,3,3,3,1,3,3,3,3, 1,1,3,3,3,3,3,3,3,3, 3,3,1,3,3,3,3,1,3,3, 3,3,1,1,3,1,3,1,1,3, 3,1,1,3,3,3,3,3,3,3, 3,3,3,3,3,3,3,1,3,3\\
       1662,\ldots, 1721& 3,3,1,3,3,3,3,3,3,3, 3,3,1,3,3,3,1,3,3,3, 3,3,1,3,3,3,3,3,3,3, 3,3,3,3,3,3,3,3,3,3, 3,3,3,3,3,3,3,3,3,3, 3,3,3,3,3,3,3,3,3,3\\
       \hline
       1722,\ldots, 1751 & 1,1,1,1,1,1,1,1,1,1, 1,1,1,1,1,1,1,1,1,1, 1,1,1,1,1,1,1,1,1,1\\
       \hline
\end{tabular}
\caption{List of divisor for each algebraic cycle.}
\label{table:divisor of alg cycles}
\end{table}
\end{theorem}

\section{Constructing Fluxes from Algebraic Cycles}\label{sect:alg_cycles}
Let us recall that in the Fermat sextic fourfold, the G-flux must be a primitive class $G\in H^{2,2}(X)_\prim$ satisfying the quantization condition 
\begin{equation}
\label{eqquantcond}
G+\frac{H^2}{2}\in H^4(X,\Z),
\end{equation}
the tadpole bound
\begin{equation}
\label{eqtadpolebound}
G\cdot G\le 217.5\, ,
\end{equation}
and has to be a general Hodge cycle \eqref{eq:genHodge}. In order to search for candidates for the G-flux we will do some approximations. 

In view of the fact that $\frac{1}{2}H^2$ has a small norm with respect to the intersection product, we will first look for a general Hodge cycle $\widetilde{G}\in H^4(X,\Z)_\prim$ with norm as small as possible and then we will try to define the G-flux $G$ by a slight modification of $\widetilde{G}$. 
Since we want to approach this using algebraic cycles, we will look for such cycles in the sub-lattice 
\begin{equation}\label{eq:sublattice}
\Gamma:=\langle \texttt{Alg}\rangle \cap H^4(X,\Z)_\prim.
\end{equation}
As explained in Section \ref{subsect:Hodgeloci} in order to check that a given Hodge cycle $\widetilde{G}$ is general, it is enough to verify that the matrix $\rho(\widetilde{G})$ defined in \eqref{eq:rhomatrix} has maximal rank, $426$. 
As $\rho_{IJ}$ is only sensitive to the primitive part of $G$, this implies that 
$\rho(G)$ has rank $426$ as well. With these considerations the problem boils down to the following optimization problem:
\begin{align}
\label{eqoptprob}    
&\min\left\{\widetilde{G}^2 \ | \ \widetilde{G}=\sum_{i=1}^{1751}n_i\cdot(A_i-d_i\cdot C_0) \ , \ n_i\in \Z\right\} \\
&\text{subject to: } \hspace{1cm} \text{rank }\rho(\widetilde{G})=426
\end{align}
where $\texttt{Alg}=\{A_i\}_{i=1}^{1751}$, $C_0$ is any linear cycle not contained in $\texttt{Alg}$ and $d_i=\deg(A_i)$ (which is 1 for a linear cycle, 6 for an Aoki-Shioda cycle and 12 for a type 3 cycle).

\subsection{General primitive algebraic cycles with small norm}

The problem \eqref{eqoptprob} is hard both theoretical and computationally. If one does a brute force searching trying to maximize the rank of $\rho(\widetilde{G})$, the experiments frequently give combinations of the form 
\begin{equation}
\label{eqdiffalgcycsametype}
\widetilde{G}=\sum_{i_1,i_2=1}^{1001}\ell_{i_1,i_2}(C_{i_1}-C_{i_2})+\sum_{j_1,j_2=1}^{720}m_{j_1,j_2}(S_{j_1}-S_{j_2})+\sum_{k_1,k_2=1}^{30}n_{k_1,k_2}(T_{k_1}-T_{k_2}).
\end{equation}
Theoretically this reduction has the advantage of preserving the orthogonality of algebraic cycles of different type, which is not preserved if we consider instead the translated algebraic cycles $A_i-d_i\cdot C_0$. Moreover, if we restrict our attention to algebraic cycles $A_i-A_j$ where $A_i$ and $A_j$ are cycles of the same type, then the values of smallest norm for linear, Aoki-Shioda and type 3 cycles are $40=2\left(\frac{125}{6}-\frac{5}{6}\right)$, $60=2(60-30)$, and $96=2(96-48)$, respectively, which are smaller than those of $A_i-d_i\cdot C_0$ for each type. In general, the values of the intersection between the primitive parts of cycles of the same type
are
\begin{equation*}
  \langle A_i, A_j\rangle_{\prim}= \left\{
\begin{array}{ll}
-\frac{25}{6},-\frac{1}{6},\frac{5}{6},\frac{125}{6}  &A_i, A_j\text{ are linear cycles},\\ 
0,\pm 3, \pm 6, \pm 12, \pm 15, \pm 30,  60   &A_i, A_j\text{ are Aoki-Shioda cycles},\\ 
0, \pm 6, \pm 12, \pm 24, \pm 48, 96  &A_i, A_j\text{ are type 3 cycles}.\\
\end{array}\right.
\end{equation*}
Note that the self-intersection for linear, Aoki-Shioda and type 3 cycles are $\frac{125}{6}, 60, 96$, respectively.

Using the reduction \eqref{eqdiffalgcycsametype} we can always consider the integers $\ell_{i_1,i_2}$,  $m_{j_1,j_2}$ and $n_{k_1,k_2}$ to be non-negative. Furthermore, noting that the intersection of two algebraic cycles of the same type is smaller than their norms, one can suppose that the size of the norm of $\widetilde{G}$ is correlated with the sum of the norms of each difference. Hence a coarse linear approximation of $\widetilde{G}^2$ is the function 
$$
\varphi(\ell,m,n):=40\ell+60m+96n
$$
where $\ell$ is the amount (counted with multiplicities) of differences of linear cycles in the expression \eqref{eqdiffalgcycsametype} of $\widetilde{G}$, $m$ is the amount of differences of Aoki-Shioda cycles and $n$ is the amount of differences of type 3 cycles.

In order to linearize the restriction in the optimization problem \eqref{eqoptprob}, we can first approximate naively
$$
\text{rank}(\rho(\widetilde{G}))\approx \sum_i \text{rank}(\rho(A_i)),
$$
for $\widetilde{G}=\sum_{i} n_i\cdot A_i$, where each $n_i\neq 0$. Depending on the type of algebraic cycle we have  that the rank of the matrix $\rho(A_i)$ is 
\begin{equation*}
    \text{rank}(\rho(A_i))=
   \left\{
\begin{array}{ll}
19 &A_i\text{ is a linear cycle},\\ 
62 &A_i\text{ is a Aoki-Shioda  cycle},\\
106 &A_i\text{ is a type 3 cycle}.\\
\end{array}\right.
\end{equation*}
With all these considerations in mind we pose our first linear approximation to \eqref{eqoptprob} as follows
\begin{align}
\label{eq1stlinoptprob}
&\min \left\{\varphi(\ell,m,n)=40\ell+60m+96n \ | \ \ell,m,n\in \Z_{\ge 0}\right\} \\
&\text{subject to: } \hspace{1cm} 2(19\ell+62m+106n)\ge 426.
\end{align}
The solution of this problem is $\min (\varphi)=216$ and is attained for $(\ell,m,n)=(0,2,1)$, which is quite close to the tadpole bound \eqref{eqtadpolebound}. This suggests one has to search for combinations given as a sum of two differences of Aoki-Shioda cycles and one difference of type 3 cycles. However, after computational research one gets combinations of small norm but with $\text{rank}(\rho(\widetilde{G}))<426$. Hence, a better linear approximation to the rank of $\rho(\widetilde{G})$ is needed. After several computations we summarize the behavior of the ranks we found for differences of two Aoki-Shioda cycles and two type 3 cycles in the following table:

\begin{table}[!ht]
\begin{footnotesize}
\centering
\begin{tabular}{|l|l|l|l|}
\hline
Type&Hodge cycle&Intersection values& Rank\\
\hline
\hline
Linear & $\widetilde{G}=C_i-C_j$ & $\langle C_i, C_j\rangle_{\prim}=5/6$ & $\text{rank}(\rho(\widetilde{G}))\leq 38$\\
\hline
Aoki-Shioda & $\widetilde{G}=S_i-S_j$ & $\langle S_i, S_j\rangle_{\prim}=30$ & $\text{rank}(\rho(\widetilde{G}))\leq 62$\\
\hline
Aoki-Shioda & $\widetilde{G}=S_i-S_j$ & $\langle S_i, S_j\rangle_{\prim}=15$ & $\text{rank}(\rho(\widetilde{G}))\leq 80$\\

\hline
Aoki-Shioda & $\widetilde{G}=S_i-S_j$ & $\langle S_i, S_j\rangle_{\prim}=12$ & $\text{rank}(\rho(\widetilde{G}))\leq 106$\\

\hline
Type 3 & $\widetilde{G}=T_i-T_j$ & $\langle T_i, T_j\rangle_{\prim}=48$ & $\text{rank}(\rho(\widetilde{G}))\leq 106$\\
\hline
Type 3 & $\widetilde{G}=T_i-T_j$ & $\langle T_i, T_j\rangle_{\prim}=24$ & $\text{rank}(\rho(\widetilde{G}))\leq 178$\\
\hline
Type 3 & $\widetilde{G}=T_i-T_j$ & $\langle T_i, T_j\rangle_{\prim}=12$ & $\text{rank}(\rho(\widetilde{G}))\leq 210$\\

\hline
\end{tabular}
\caption{Constraints on the ranks of $\rho(\widetilde{G})$ for differences of algebraic cycles.}
\label{table:restriction on ranks}
\end{footnotesize}
\end{table}

With these restrictions, we can pose the following linear optimization problem
\begin{align}\label{eq:lop}
&\min \{\varphi=40\ell+60m_1+90m_2+96m_3+96n_1+144n_2+168n_3 \ | \ \ell,n_i,m_j\in\Z_{\ge 0}\} \\
&\text{subject to: }\hspace{5mm} R=38\ell+62m_1+80m_2+106m_3+106n_1+178n_2+210n_3\ge 426.
\end{align}
Below is the solution to this optimization problem, and also the next smallest possible values attained
\begin{itemize}
    \item $(\ell,m_1,m_2,m_3,n_1,n_2,n_3)=(1,0,0,0,0,1,1)$ with $\varphi=352$, and  $R=426$,
    \item $(\ell,m_1,m_2,m_3,n_1,n_2,n_3)=(2,0,0,0,0,2,0)$ with $\varphi=368$, and  $R=432$,
    \item $(\ell,m_1,m_2,m_3,n_1,n_2,n_3)=(0,1,0,0,0,1,1)$ with $\varphi=372$, and  $R=450$.
\end{itemize}

To every algebraic cycle of the form 
$$
\widetilde{G}=\sum_{i=1}^\ell (C_{p_{2i-2}}-C_{p_{2i-1}})+\sum_{j=1}^m(S_{q_{2j-2}}-S_{q_{2j-1}})+\sum_{k=1}^n(T_{r_{2k-2}}-T_{r_{2k-1}})
$$
we associate an intersection matrix corresponding to the intersection matrix of the tuple of primitive algebraic cycles supporting $\widetilde{G}$, i.e. the intersection matrix of the tuple  $$
([C_{p_0}]_{\prim},\ldots,[C_{p_{2\ell-1}}]_{\prim},[S_{q_0}]_{\prim},\ldots,[S_{q_{2m-1}}]_{\prim},[T_{r_0}]_{\prim},\ldots,[T_{r_{2n-1}}]_{\prim}).
$$
These solutions to the linear optimization problem \eqref{eq:lop} would be very close to the desired solution if the intersection matrices associated to the algebraic cycle in each case were block diagonal and respectively equal to 
$$
\begin{tiny}
\begin{pmatrix}
    125/6 & 5/6  &0 &0&0 &0\\
    5/6 & 125/6  &0 &0&0 &0\\
    0 & 0  & 96 &24&0 &0\\
    0 & 0  & 24 &96&0 &0 \\
    0 & 0  &0 &0 &96 &12\\
    0 & 0  &0 &0 &12 &96\\
\end{pmatrix}
\end{tiny} , \begin{tiny}
\begin{pmatrix}
    125/6 & 5/6  & 0 &0 &0 &0&0 &0\\
    5/6 & 125/6  & 0 &0 &0 &0&0 &0\\
    0 &0& 125/6 & 5/6  & 0 &0 &0 &0\\
    0 &0&5/6 & 125/6  & 0 &0 &0 &0\\
    0 & 0  &0 &0& 96 &24 &0 &0\\
    0 & 0  &0 &0& 24 &96 &0 &0\\
    0 & 0  & 0 &0 &0 &0&96 &24\\
    0 & 0  & 0 &0 &0 &0&24 &96\\
\end{pmatrix}
\end{tiny} , \begin{tiny}
\begin{pmatrix}
    60 & 30  &0 &0&0 &0\\
    30 & 60  &0 &0&0 &0\\
    0 & 0  & 96 &24&0 &0\\
    0 & 0  & 24 &96&0 &0 \\
    0 & 0  &0 &0 &96 &12\\
    0 & 0  &0 &0 &12 &96\\
\end{pmatrix}
\end{tiny}.
$$
However this is not true, and so the norm may vary from the estimated values. Based in the solutions to the linear optimization problem \eqref{eq:lop} we computed the norm  and rank of lots of combinations of algebraic cycles for each solution, and the smallest norms found for general algebraic cycles are $312$, 316 and $336$. The following combinations of algebraic cycles are general and attain these bounds: 
\begin{itemize}
    \item $\widetilde{G}_1=(C_{994}-C_{282})+(T_{1}-T_{5})+(T_{23}-T_{18})$, with norm 316, and intersection matrix 
    \begin{equation}
%\label{Eq: Matrix_norm_324}
\begin{footnotesize}
\begin{pmatrix}
125/6 & 5/6& -4&0& 0& 0 \\    
 5/6 &125/6&-2& 2& 0& 0\\        
 -4&-2&96& 24&-6&6\\        
 0& 2&24&96& 0&-6\\
 0& 0&-6&0&96& 12\\        
 0& 0& 6&-6&12&96
\end{pmatrix},
\end{footnotesize}
\end{equation}

\item $\widetilde{G}_2=(C_{232}-C_{675})+(T_{4}-T_{2})+(T_{19}-T_{21})$, with norm 316, and intersection matrix 
    \begin{equation}
%\label{Eq: Matrix_norm_324}
\begin{footnotesize}
\begin{pmatrix}
125/6 & -1/6& 0&4& 0& 0 \\    
 -1/6 &125/6&0& 0& -1& 0\\        0&0&96&24&-6&0\\        
 4&0&24&96&6&-6\\
 0&-1&-6&6&96&12\\        
 0&0&0&-6&12&96\\        
     
\end{pmatrix},
\end{footnotesize}
\end{equation}

\item $\widetilde{G}_3=(C_{841}-C_{923})+(C_{660}-C_{762})+(T_{8}-T_{3})+(T_{24}-T_{28})$, with norm 312 and intersection matrix 
\begin{equation}
\label{Eq: Matrix_norm_332}
\begin{footnotesize}
\begin{pmatrix}
125/6&-1/6&-1/6&-1/6&-1&1&-1&0\\
-1/6&125/6&-25/6&5/6&-2&-2&0&0\\
-1/6&-25/6&125/6&-1/6&2&4&0&2\\
-1/6&5/6&-1/6&125/6&0&0&4&0\\
-1&-2&2&0&96&24&0&0\\
1&-2&4&0&24&96&24&-12\\
-1&0&0&4&0&24&96&12\\
0&0&2&0&0&-12&12&96
\end{pmatrix},
\end{footnotesize}
\end{equation}
%The intersection matrix \eqref{Eq: Matrix_norm_332} in the normalized basis is 
%\begin{equation}\label{Eq: Matrix_norm_332_normalized}   \begin{footnotesize}        \begin{pmatrix}            42&   22&  248&  252&  252&  252&\\22&   42&  250&  252&  252&  254\\248&  250& 3096& 3006& 2994& 3024\\252&  252& 3006& 3096& 3012& 2994\\252&  252& 2994& 3012& 3096& 3000\\252&  254& 3024& 2994& 3000& 3096        \end{pmatrix}    \end{footnotesize}\end{equation}

\item $\widetilde{G}_4=(C_{307}-C_{614})+(C_{280}-C_{421})+(T_{3}-T_{8})+(T_{11}-T_{16})$, with norm 312 and intersection matrix \begin{equation}
\label{Eq: Matrix_norm_332_v1}
\begin{footnotesize}
\begin{pmatrix}
125/6&-1/6&-1/6&-1/6& 0&0&-2&4\\       
-1/6&125/6& 5/6&-1/6& 0&0&2&-4\\       
-1/6& 5/6&125/6& 5/6&-2&2&0&0\\       
-1/6&-1/6& 5/6&125/6&2&-2&1&-1\\       
0&0&-2&2&96&24&0&6\\       
0&0&2&-2&24&96&6&6\\       
-2&2&0&1&0&6&96&24\\       
4&-4&0&-1&6&6&24&96
\end{pmatrix} ,
\end{footnotesize}
\end{equation}

\item $\widetilde{G}_5=(S_{671}-S_{505})+(T_{27}-T_{3})+(T_{5}-T_{7})$, %1671, 1505, 1747, 1723, 1725, 172
with norm 336 and intersection matrix \begin{equation}
\label{Eq: Matrix_norm_336}
\begin{footnotesize}
\begin{pmatrix}
60&   0&   0&   0&   0&   0\\
0&  60&   0&   0&   0&   0\\
0&   0&  96&  12& -12&   0\\
0&   0&  12&  96&   0& -48\\
0&   0& -12&   0&  96&  12\\
0&   0&   0& -48&  12&  96
\end{pmatrix} .
\end{footnotesize}
\end{equation}

%\item $G_6=(S_{525}-S_{104})+(T_{7}-T_{5})+(T_{3}-T_{27})$, 
%[1525, 1104, 1727,1725 , 1723, 1747]
%with norm 336 and intersection matrix \begin{equation}\label{Eq: Matrix_norm_336_v1}\begin{footnotesize}\begin{pmatrix}60&   0&  0&   0&   0&   0\\0&  60&   0&   0&   0&   0\\0&   0&  96&  12& -48&   0\\0&   0&  12&  96&   0& -12\\0&   0& -48&   0&  96&  12\\0&   0&   0& -12&  12&  96\end{pmatrix} .\end{footnotesize}\end{equation}

%The intersection matrix \eqref{Eq: Matrix_norm_332_v1} in the normalized basis is \begin{equation}   \label{Eq: Matrix_norm_332_normalized_v1}\begin{footnotesize}\begin{pmatrix}      42&  21&  21&  20& 252& 252& 250& 256\\             21&  42&  22&  20& 252& 252& 254& 248\\             21&  22&  42&  21& 250& 254& 252& 252\\             20&  20&  21&  40& 242& 238& 241& 239\\             252& 252& 250& 242&3096&3024&3000&3006\\         252& 252& 254& 238&3024&3096&3006&3006\\        250& 254& 252& 241&3000&3006&3096&3024\\         256& 248& 252& 239&3006&3006&3024&3096\end{pmatrix}\end{footnotesize}\end{equation}
\end{itemize}

\subsection{Constructing a G-flux}

Now we want to construct a candidate for the G-flux from each primitive algebraic cycle with small norm found in the previous section. The only issue here is the quantization condition \eqref{eqquantcond} which corresponds to belonging to
$$
\Lambda_{phys}=\left\{G\in H^{2,2}(X)_\prim | G+\frac{H^2}{2}\in H^4(X,\Z)\right\},
$$
which is not a lattice. In spite of this, if we pick any $G_0\in \Lambda_{phys}$ then the translation
$$
\Lambda_{phys}-G_0=H^{2,2}(X)_\prim\cap H^4(X,\Z)
$$
is the lattice of primitive integral Hodge cycles. Hence our candidate for G-flux will be of the form 
$
\widetilde{G}+G_0,
$
where $\widetilde{G}=\widetilde{G}_i$ for $i=1,2,3,4,5$ are the primitive algebraic cycles found in the previous section. Since we want the translation not to change the norm of $\widetilde{G}$ too much, we ask first which elements of $\Lambda_{phys}$ have the smallest norm. Since $$G_0=(G_0+\frac{H^2}{2})_\prim$$ and $$G_0+\frac{H^2}{2}\in H^{2,2}(X)\cap H^4(X,\Z),$$ in view of the integral Hodge conjecture we can suppose $G_0$ is the primitive part of an algebraic cycle $$L=G_0+\frac{H^2}{2}$$ whose class in $\P^5$ is $\frac{H^2}{2}\cdot 6H=3H^3$, i.e. a cycle of degree 3. In view of the three types of algebraic cycles we have at our disposal in the Fermat variety, the minimal way to produce such a degree 3 cycle is by taking $$L=C_i+C_j+C_k$$ for three linear cycles. Hence, our candidate for the G-flux will be of the form
$$
G=\widetilde{G}+(C_i+C_j+C_k)_\prim.
$$
For each of the cycles $\widetilde{G}\in\{\widetilde{G}_1,\widetilde{G}_2,\widetilde{G}_3,\widetilde{G}_4,\widetilde{G}_5\}$ of the previous section, we find experimentally that the $G$ with lowest  norm is obtained when one of the three linear cycles appears with a negative sign in the expression of $\widetilde{G}$. More precisely, for $\widetilde{G}=\widetilde{G}_1$, we get
\begin{equation}
\label{eqGflux1}
G^2=(\widetilde{G}_1+(C_{231}+C_{282}+C_{562})_\prim)^2=293.5.
\end{equation}
For $\widetilde{G}=\widetilde{G}_2$, we get 
\begin{equation}
\label{eqGflux2}
G^2=(\widetilde{G}_2+(C_{12}+C_{14}+C_{675})_\prim)^2=299.5.   
\end{equation}
For $\widetilde{G}=\widetilde{G}_3$, we get 
\begin{equation}
\label{eqGflux3}
G^2=(\widetilde{G}_3+(C_{169}+C_{595}+C_{923})_\prim)^2=293.5.   
\end{equation}
For $\widetilde{G}=\widetilde{G}_4$, we get 
\begin{equation}
\label{eqGflux4}
G^2=(\widetilde{G}_4+(C_{55}+C_{101}+C_{614})_\prim)^2=315.5.   
\end{equation}
For $\widetilde{G}=\widetilde{G}_5$, we perform the calculation for various sums of three linear cycles, and the combination with the lowest norm that we obtain is
\begin{equation}
\label{eqGflux5}
G^2=(\widetilde{G}_5+(C_{901}+C_{914}+C_{871})_\prim)^2=353.5.   
\end{equation}
These candidates for the G-flux \eqref{eqGflux1}, \eqref{eqGflux2}, \eqref{eqGflux3}, \eqref{eqGflux4} and \eqref{eqGflux5}  are such that $\rho(G)$ is of full rank.

\section{Constructing Fluxes from Residues}\label{sect:residues}
In this section we will approach the problem of finding a Hodge cycle of maximal condimension and minimal length by working with the residues $\omega_\beta$. For any primitive Hodge cycle $G$ we can write 
\begin{equation}\label{eq:defG_exp_residues}
G = \sum_{\beta, |\beta|=2} \alpha_\beta\,  \omega_\beta \, .
\end{equation}

The components of $\rho_{IJ}$ can be worked out using \eqref{eq:inner_form_residues}: 
\begin{equation}\label{rho_entries}
\rho_{IJ} =  \sum_{|\beta|=2} \alpha_\beta \frac{- (2 \pi i)^4}{2^2 6^5} \omega_{\beta_I + \beta_J} \cdot \omega_\beta  =\frac{- (2 \pi i)^4}{2^7 3^5} \alpha_{\,\overline{\beta_I+\beta_J}}\,  \\
\end{equation}
For ease of computation, we rescale $\rho_{IJ}$ by the global prefactor and instead work with
\begin{equation}
\widetilde{\rho_{IJ}} \equiv  \frac{-2^7 3^5}{(2 \pi i)^4} \rho_{IJ}
= \alpha_{\,\overline{\beta_I+\beta_J}}
\end{equation}

Given that $H^{4}(X,\Z)$ is a self-dual lattice and that we know a basis of $H^{4}(X,\Z)$ in terms of vanishing cycles and linear algebraic cycles, as well as the intersection between the residues and any vanishing cycle or algebraic cycle using \eqref{eq:inner_form_residues} and \eqref{eq:period_vancyc_residue}, we have that $G + c_2(X)/2 \in H^{4}(X,\Z)$ if and only if 
\begin{equation}\label{eq:int_van_cycle_flux}
G \cdot \delta_\beta \in \Z \,\, \forall \beta
\end{equation}
and 
\begin{equation}\label{eq:int_alg_cycle_flux}
\left( G + \frac{15 H^2}{2} \right)\cdot C^{\ell}_\sigma = G\cdot C^{\ell}_\sigma + \frac{15}{2} \in \Z \,\, \mbox{for some} \,\, C^{\ell}_\sigma\, .
\end{equation}
By imposing these conditions on the Ansatz \eqref{eq:defG_exp_residues}, which is primitive and of Hodge type $(2,2)$ by construction, we can hence find the set $\Lambda_{phys}$ of all properly quantized Hodge cycles expressed in terms of residues. For any $G \in \Lambda_{phys}$ we can then straightforwardly work out $\widetilde{\rho_{IJ}}$ and $G\cdot G$. 

In principle, this then allows us to use the following strategy: for any bound $T$, we can  
list all $G \in \Lambda_{phys}$ with $G\cdot G \leq T$, and then check what is the maximal rank of $\rho$ found within this set. If we can perform this task for choices of $T$ that are big enough, this in turn allows us to determine the minimal value of $G\cdot G$ for which $\rho$ has maximal rank. 
As $\Lambda_{phys}$ has rank $1751$ this is a formidable problem that is computationally beyond our present capabilities. 

We will hence content ourselves with addressing a somewhat simplified problem in this section and study this problem for Fermat sextics with specific symmetries. In other words, we will select a finite abelian groups of symmetries $\Gamma$ of $X$ and only consider fluxes $G$ and deformations invariant under the action of $\Gamma$. In practice, this achieved by restricting the $\omega_\beta$ appearing in $G$ and those appearing in $\rho_{IJ}$ to monomials invariant under $\Gamma$. Besides giving us tractable versions of the question we are interested in for the Fermat sextic $X$, we also get flux solutions for the quotients $X/\Gamma$ (almost) for free. 

Before discussing several examples of $\Gamma$ in detail, let us collect a few useful facts and establish some notation. 

\subsection{Complex Conjugation}\label{CC}

As we want $G$ in particular to be real, $G \in H^4(X)$, we need to study the action of complex conjugation. As $\omega_\beta$ is primitive and a basis of the primitive middle cohomology is given by vanishing cycles tensored with $\R$, it follows from \eqref{eq:period_vancyc_residue} that $\bar{\omega}_{\beta}$ is proportional to $\omega_{\bar{\beta}}$. We set
\begin{equation}
\bar{\omega}_{\beta}  = c_\beta \omega_{\bar{\beta}}\, .
\end{equation}
We can now fix $c_\beta$ by integrating both $\bar{\omega}_\beta$ and $\omega_{\bar{\beta}}$ over the same vanishing cycle. As this cycle is real
\begin{equation}
\int_{\delta_{\beta'}} \bar{\omega}_\beta = \overline{\int_{\delta_{\beta'}} \omega_\beta}\, .
\end{equation}
Using $\beta' = (0^6)$ we find for $|\beta|=2$ that
\begin{equation}
c_\beta = \frac{ \overline{\int_{\delta_{\beta'}} \omega_\beta}}{ \int_{\delta_{\beta'}} \omega_{\bar{\beta}}} =
-\prod_{i=0}^5\frac{\Gamma\left(\frac{\beta_i+1}{6}\right)}{\Gamma\left(\frac{5-\beta_i}{6}\right)}\frac{\overline{\zeta_6^{\beta_i+1}-1}}{\zeta_6^{5-\beta_i}-1}
=-\prod_{i=0}^5\frac{\Gamma\left(\frac{\beta_i+1}{6}\right)}{\Gamma\left(\frac{5-\beta_i}{6}\right)}\, .
\end{equation}
This has the structure as found above and is furthermore real in all cases. In particular, we have the following depending on decomposability of $\beta$:\\

\begin{itemize}
 \item Whenever $\beta$ is 3-decomposable we have that (choosing an appropriate permutation) $\beta_{2i} = 4-\beta_{2i+1}$, so that $c_\beta=-1$. 

 \item When $\beta$ is 1-decomposable we have $\beta = (\beta_0,4-\beta_0,0,2,3,3)$ and hence
\begin{equation}
 c_\beta = -\frac{\Gamma\left(\tfrac16\right)\Gamma\left(\tfrac12\right)\Gamma\left(\tfrac23\right)^2}{\Gamma\left(\tfrac56\right)\Gamma\left(\tfrac12\right)\Gamma\left(\tfrac13\right)^2}
 = -\frac{\Gamma\left(\tfrac16\right)\Gamma\left(\tfrac23\right)^2}{\Gamma\left(\tfrac56\right)\Gamma\left(\tfrac13\right)^2}
 = -\frac{\sqrt{\pi} 2^{1-1/3}\Gamma\left(\tfrac13\right)\Gamma\left(\tfrac23\right)}{\Gamma\left(\tfrac13\right) \Gamma\left(\tfrac23\right)\sqrt{\pi} 2^{1-2/3}} =
 -2^{1/3} \, .
\end{equation}
 \item When $\beta$ is indecomposable we can set $\beta = (4,4,1,1,1,1)$ and hence 
 \begin{equation}
 c_\beta = -\frac{\Gamma\left(\tfrac56\right)^2\Gamma\left(\tfrac13\right)^4}{\Gamma\left(\tfrac16\right)^2\Gamma\left(\tfrac23\right)^4} 
 =-\frac{\Gamma\left(\tfrac23\right)^2 \left(2^{1-2/3}\sqrt{\pi} \right)^2 2\Gamma\left(\tfrac13\right)^2}{\Gamma\left(\tfrac23\right)^2 \left(2^{1-2/6}\sqrt{\pi} \right)^2 2\Gamma\left(\tfrac13\right)^2}
 = -\frac{2^{2/3}}{2^{4/3}} =-2^{-2/3} \, .
 \end{equation}
\end{itemize}

Here we have used the following relations repeatedly
\begin{equation}
\begin{aligned}
\Gamma(z)\Gamma(1-z) &= \frac{\pi}{\sin(\pi z)} \\ 
\Gamma(z)\Gamma(z+\tfrac12) &= 2^{1-2z} \sqrt{\pi}\, \Gamma(2z) \, .
\end{aligned} 
\end{equation}

\subsection{The period formula, Integrality, and $G^2$}

In order to efficiently impose \eqref{eq:int_van_cycle_flux} we introduce some suitable notation. We can write \eqref{eq:period_vancyc_residue} for $|\beta|=2$ as
\begin{equation}
\begin{aligned}
\int_{\delta_{\beta'}} \omega_\beta &= \frac{1}{6^5} \frac{1}{4\pi i} \prod_i \Gamma\left(\frac{\beta_i+1}{6}\right) \left(\zeta_6^{(\beta_i'+1)(\beta_i+1)}
-\zeta_6^{(\beta_i')(\beta_i+1)} \right)  \\
& = \frac{1}{6^5} \frac{1}{4\pi i} \prod_i \Gamma\left(\frac{\beta_i+1}{6}\right) \left(\zeta_6^{(\beta_i+1)}-1 \right)\zeta_6^{\beta_i'(\beta_i+1)} \\
& = z_u z_\beta Z(\beta,\beta')
\end{aligned}
\end{equation}
with
\begin{equation}
\begin{aligned}
z_u & = \frac{1}{6^5} \frac{1}{4\pi i}\\
 z_\beta &=  \prod_i \Gamma\left(\frac{\beta_i+1}{6}\right)\left(\zeta_6^{(\beta_i+1)}-1 \right)\\
 Z(\beta,\beta') & = \zeta_6^{\sum_i \beta_i'(\beta_i+1)} 
 \end{aligned}\,.
\end{equation}
Here, $z_u$ is a universal normalization factor and the factors $z_\beta$ are a normalization which only depends on $\beta$, i.e. on the residue in question. The only non-trivial data about the lattice determined by \eqref{eq:int_van_cycle_flux} is contained in $Z(\beta,\beta')$ which are always some sixth root of unity for all $\beta$ or $\beta'$. Note that
\begin{equation}
Z(\bar{\beta},\beta') = \zeta_6^{\sum_i \beta_i'(\bar{\beta}_i+1)}
= \zeta_6^{\sum_i \beta_i'(5-\beta_i)} =\overline{ \zeta_6^{\sum_i \beta_i'(\beta_i+1)} }
= \overline{Z(\beta,\beta')}
\end{equation}

We now make the following manifestly real Ansatz for $G$:
\begin{equation}\label{eq:G_normalized}
G = \sum_{\beta \in \mathcal{I}} \frac{\nu_\beta}{z_u z_\beta}\,  \omega_\beta +  \frac{\bar{\nu}_\beta}{\bar{z}_u \bar{z}_\beta}\,  \bar{\omega}_\beta \, .
\end{equation}
Here, $\mathcal{I}$ a subset of $\beta$s with $|\beta|=2$ which contains $\beta = (2^6)$ and exactly one member from each pair $\beta,\bar{\beta}$. This set has $876$ elements. Note that $\bar{\omega}_{(2^6)}=-\omega_{(2^6)}$, $\bar{z}_{(2^6)} =  z_{(2^6)}$ and $\bar{z}_u = - z_u$ which implies that $\nu_{(2^6)}$ is real. This implies that 
\begin{equation}
\begin{aligned}
\alpha_\beta &= \frac{\nu_\beta}{z_u z_\beta} \hspace{1cm} \mbox{if}\,\,\, \beta \in \mathcal{I} \setminus (2^6)\\
\alpha_\beta &= \frac{\bar{\nu}_{\bar{\beta}}c_{\bar{\beta}}}{\bar{z}_u \bar{z}_{\bar{\beta}}} \hspace{1cm} \mbox{if}\,\,\, \beta \notin \mathcal{I} \\
\alpha_{(2^6)} &= \frac{2\nu_{(2^6)}}{z_u z_{(2^6)}}
\end{aligned}
\end{equation}

Using this Ansatz, condition \eqref{eq:int_van_cycle_flux} becomes 
\begin{equation}\label{eqZZ*condition}
    G \cdot \delta_{\beta'} = 
    \sum_{\beta \in \mathcal{I}} \nu_\beta Z(\beta,\beta') +  \bar{\nu}_\beta \bar{Z}(\beta,\beta') \in \Z \,\, \forall \beta'
\end{equation}
Defining the inner form
\begin{equation}
\langle a ,b \rangle \equiv a \bar{b} + \bar{a} b
\end{equation}
on $\C$, the above relations can now be understood as the condition that the $\nu_\beta$ are contained in the lattice $\mathcal{Z}^*$ dual to the lattice $\mathcal{Z}$ spanned by the $Z(\beta,\beta')$.  

Note that with the inner form just defined, the lattice generated by sixth roots of unity is isometric to the root lattice $A_2$, and the lattice generated by $1$ becomes isometric to the root lattice $A_1$. 

We can make a few general statements about the lattice $\mathcal{Z}$ by examining the behavior of the $Z(\beta,\beta')$. If $\beta = (2,2,2,2,2,2)$, $Z(\beta,\beta')$ is always $\pm 1$. For all other $\beta$, there is always a $\beta'$ such that $Z(\beta,\beta')$ becomes a non-trivial third or sixth root of unity.  Furthermore, as $\beta' = (0,0,0,0,0,0)$ appears as a condition in \eqref{eqZZ*condition}, $Z(\beta,\beta') =1$ also appears for every $\beta$. This means that using the inner form above for each fixed $\beta$, $Z(\beta,\beta')$ is contained in $A_1$ if $\beta = (2,2,2,2,2,2)$, and is contained in $A_2$ for all others. We hence have that\footnote{We shall see in examples below that $\mathcal{Z} \neq  A_2^{875} \oplus A_1$, so we have in fact a proper inclusion. 
} 
\begin{equation}\label{eq:latticeZinclusion}
\mathcal{Z} \equiv   Span_\Z (\{ \oplus_{\beta \in \mathcal{I}} Z(\beta,\beta')\} )  \subseteq A_2^{875} \oplus A_1 \, .
\end{equation}
Note that $\bar{\nu}_\beta \in \mathcal{Z}^*$ alone does not yet imply that $G \in \Lambda_{phys}$, as we also need to satisfy \eqref{eq:int_alg_cycle_flux}.

Having determined $\Lambda_{phys}$ by appropriately chosing the $\nu_\beta$ we can work out 
\begin{equation}
\begin{aligned}
G \cdot G &= \frac{4 \nu_{(2^6)}^2}{|z_u|^2 z_{(2^6)}^2} \omega_{(2^6)} \cdot \bar{\omega}_{(2^6)}
+ \sum_{\beta \in \mathcal{I}\setminus (2^6)}  \frac{ 2\nu_\beta\bar{\nu}_\beta}{|z_u|^2|z_\beta|^2}\omega_{\beta} \cdot \bar{\omega}_{\beta} \\
&= (-1) \frac{(2\pi i)^4}{2^2} \frac{1}{6^5} 6^{10} |4 \pi i|^2 \left[-\frac{4 \nu_{(2^6)^2}}{z_{(2^6)}^2} + \sum_{\beta \in \mathcal{I} \setminus (2^6)} \frac{2\nu_\beta\bar{\nu}_\beta}{|z_\beta|^2} c_\beta \right]
\end{aligned}
\end{equation}
This can be simplified using
\begin{equation}
\begin{aligned}
\frac{c_\beta}{|z_\beta|^2} &= - \prod_{i=0}^5 \Gamma\left(\frac{\beta_i+1}{6}\right)
\Gamma\left(\frac{5-\beta_i}{6}\right)^{-1} \Gamma\left(\frac{\beta_i+1}{6}\right)^{-2}
\left|\zeta_6^{\beta_i+1}-1 \right|^{-2}\\
& = \frac{1}{2^{12} \pi^6 S_\beta}
\end{aligned}
\end{equation}
with 
\begin{equation}
S_\beta \equiv \prod_{i=0}^5 \sin\left(\frac{\beta_i+1}{6}\right)
\end{equation}
Finally, $S_\beta$ can be easily worked out using
\begin{equation}
\begin{array}{c|c}
\beta_i & \sin \pi \left(\frac{\beta_i+1}{6}\right)  \\ 
\hline
0 & \tfrac12 \\
1 & \tfrac12 \sqrt{3}\\
2 & 1\\
3 & \tfrac12 \sqrt{3}\\
4 & \tfrac12
\end{array}
\end{equation}
\vspace{1cm}

\noindent
Putting it all together, we find that 
\begin{equation}\label{eq:Gsquared}
G \cdot G = 3^5 \left( \langle \nu_{(2^6)}, \nu_{(2^6)}\rangle + \frac{1}{2}\sum_{\beta \in \mathcal{I}\setminus (2^6)} \frac{\langle \nu_\beta, \nu_{\beta}\rangle}{S_\beta}\right) \, .
\end{equation}

\subsection{Groups of Symmetries}

We will consider symmetries that act on the homogeneous coordinates $x_i$ as
\begin{equation}\label{eq:symmetry_diag_action}
g: (x_0,x_1,x_2,x_3,x_4,x_5)\,\, \rightarrow \,\,
(\zeta^{g_0} x_0,\zeta^{g_1}x_1,\zeta^{g_2}x_2,\zeta^{g_3}x_3,\zeta^{g_4}x_4,\zeta^{g_5}x_5)\, ,
\end{equation}
with $\zeta$ a primitive $6$th root of unity and $g_i \in [0,1,2,3,4,5]$ , i.e. we do not consider group actions which permute the homogeneous coordinates $x_i$. We can specify any such group action by giving the weights $g_i$ of each generator of the group. Note that the homogeneous coordinates are only defined modulo the $\C^*$ action of $\P^5$, so that we can identify
\begin{equation}
 (g_0,g_1,g_2,g_3,g_4,g_5) \simeq  (g_0+1,g_1+1,g_2+1,g_3+1,g_4+1,g_5+1) \, .
\end{equation}

The action of some $g \in \Gamma$ on  $\omega_{\beta} $ is
\begin{equation}
    \omega_{\beta} \rightarrow \omega_{\beta}  \zeta^{\sum_i g_i (\beta_i+1)}
\end{equation}
so that the invariant subspace in the middle cohomology is spanned by $H^2$ together with those residues for which
\begin{equation}
\sum_i g_i (\beta_i+1)  =  0 \,\mbox{mod}\,6 \,\,\,\forall \, g \in \Gamma
\end{equation}

We will only consider groups $\Gamma$ which preserve the holomorphic top-form $\Omega$, so that we need
\begin{equation}
\sum_i g_i = 0 \,\mbox{mod}\, 6 \,\,\,\forall \, g \in \Gamma \, .
\end{equation}

We can classify all groups of such symmetries as follows. The subgroup of the automorphism group of $X$ preserving $\Omega$ and acting as \eqref{eq:symmetry_diag_action} is $\mathcal{A}_d = (\Z/6\Z)^4$ with generators 
\begin{equation}\label{eq:weightsz64}
\begin{array}{cccccc}
g_0 & g_1 & g_2 & g_3 & g_4 & g_5 \\
1 & -1 &  &  &  &  \\
 & 1 & -1 &  &  &  \\
 &  & 1 & -1 &  &  \\
 &  &  & 1 & -1 &  \\
\end{array}
\,\,\simeq\,\,
\begin{array}{cccccc}
g_0 & g_1 & g_2 & g_3 & g_4 & g_5 \\
1 &  &  &  & -1 &  \\
 & 1 &  &  & -1 &  \\
 &  & 1 &  & -1 &  \\
 &  &  & 1 & -1 &  \\
\end{array}
\end{equation}
Note that this group action does not single out any homogeneous coordinate despite its presentation. Any subgroup $\Gamma \triangleleft \mathcal{A}_d$ hence gives us an instance of a faithful representation $\Gamma$ that acts diagonally as \eqref{eq:symmetry_diag_action} and preserves the Calabi-Yau property. Conversely, consider the image of an element $g \in \Gamma$ under a representation $r$ that acts on $X$ as \eqref{eq:symmetry_diag_action} and preserves the Calabi-Yau property. It follows that $r(g) \in \mathcal{A}_d$, so that $r(\Gamma)$ must be a subgroup of $\mathcal{A}_d$. If $r$ is furthermore faithful we have that $r(\Gamma) \cong \Gamma$. 

We can hence classify all finite abelian groups which faithfully act in a representation of the form \eqref{eq:symmetry_diag_action} on $X$ and which preserve the Calabi-Yau property by finding all sungroups of $\mathcal{A}_d$. There are $14204$ subgroups of $\mathcal{A}_d$, however many of these are identified when permuting the coordinates $x_i$. In such cases, the dimensions of the invariant subspaces of $H^{p,q}$ are equal. We have listed all possible orders of $\Gamma$ together with the dimension of the invariant subspaces of $H^{3,1}(X)$ and $H^{2,2}(X)$ in Table \ref{tab:possible_groups}. Note that a single entry potentially corresponds to genuinely different subgroups of $(\Z/6\Z)^4$, i.e. subgroups which are not identified by simply permuting the $x_i$.

\subsection{Symmetric under $\Gamma= \left(\mathbb{Z}/6\mathbb{Z}\right)^4$}\label{sect:Z6to4}

In this example, we define the action of the group $\Gamma$ by
\begin{equation}
\begin{array}{cccccc}
g_0 & g_1 & g_2 & g_3 & g_4 & g_5 \\
1 & -1 & 0 & 0 & 0 & 0 \\
0 & 1 & -1 & 0 & 0 & 0 \\
0 & 0 & 1 & -1 & 0 & 0 \\
0 & 0 & 0 & 1 & -1 & 0 \\
\end{array}\, .
\end{equation}
For a form $\omega_\beta$ to be invariant we need that $\beta_i = \beta_j$ for all $i,j$. The only invariant residues are hence
\begin{equation}
\begin{array}{ccc}
& \beta & |\beta| \\
\hline
\omega_{(0^6)} & (0,0,0,0,0,0) & 0 \\ 
\omega_{(1^6)} & (1,1,1,1,1,1) & 1 \\ 
\omega_{(2^6)} & (2,2,2,2,2,2) & 2 \\ 
\omega_{(3^6)} & (3,3,3,3,3,3) & 3 \\ 
\omega_{(4^6)} & (4,4,4,4,4,4) & 4 \\ 
\end{array}
\end{equation}
In particular, there is now only a single term in $G$ that is non-zero and the matrix $\rho$ is just a number. 

Imposing that $G \cdot \delta_{\beta'} \in \Z$ now results in $\bar{\nu}_{(2^6)}  \in A_1^*$ which implies that $\nu_{(2^6)} = \frac{n}{2}$ for $n \in \Z$ so that 
\begin{equation}
G = n\frac{\omega_{(2^6)}}{z_u z_{(2^6)}}    =
n \frac{ 2 \,\cdot \, 3^5 \, i}{\pi^2}\,\,  \omega_{(2^6)}
\end{equation}
Next we impose that 
\begin{equation}
G \cdot C_\ell \in \Z + \frac{1}{2} \, ,
\end{equation}
which for $C_{0,0,0}$ reads 
\begin{equation}
\int_{C_{id}^{0,0,0}} G = n \frac{ 2 \,\cdot \, 3^5 \, i}{\pi^2}\,\,  (2\pi i)^2 \frac{1}{2 \cdot 6^3} \zeta_{12}^{\sum_{e=1}^3 (1+2)1} 
= -n \frac{3^2}{2}\in \Z + \frac{1}{2}
\end{equation}
so that $n$ must be odd, $n = 2m+1$ for $m \in \Z$. 

We hence find that $\Lambda_{phys}$ is described as 
\begin{equation}
    \Lambda_{phys} = \left\{ G_m = (2m+1) \frac{ 2 \,\cdot \, 3^5 \, i}{\pi^2}\,\,  \omega_{(2^6)} | m \in \Z \right\}\,.
\end{equation}
For any element of $\Lambda_{phys}$ we have
\begin{equation}
\begin{aligned}
G_m^2 & = \left( (2m+1) \frac{ 2 \,\cdot \, 3^5 \, i}{\pi^2} \right)^2\,\,  \omega_{(2^6)}\cdot \omega_{(2^6)} =
-(2m+1)^2 \frac{2^2 \cdot 3^{10}}{\pi^4} \left(-\frac{(2 \pi i)^4}{2^7 \cdot 3^5}\right) \\
& = \frac{(2m+1)^2 \, 3^5}{2}
\end{aligned}
\end{equation}

The matrix $\rho$ is just a number in this case and it is non-zero, i.e. has full rank, whenever
$G \neq 0$, which is true for any $m$. The shortest choices of $G$ are $m=0$ and $m=-1$ for which 
\begin{equation}
G^2_{0} = G^2_{-1} = \frac{3^5}{2} \, .
\end{equation}
Note that 
\begin{equation}
\frac{\chi(X)}{24} -\frac{1}{2} G^2_{0} = 48 > 0 \,,
\end{equation}
so that this flux is a perfectly viable solution. However, choosing the next to shortest flux for $m=1$ results in 
\begin{equation}
\frac{\chi(X)}{24} -\frac{1}{2} G^2_{1} = -438 < 0 
\end{equation}
and does hence not give a consistent solution.
\\

\subsection{Symmetric under $\Gamma= \left(\Z/6\Z\right)^3 \times \left(\Z/3\Z\right)$}

Here, $\Gamma$ is defined by its generators acting as
\begin{equation}
\begin{array}{cccccc}
g_0 & g_1 & g_2 & g_3 & g_4 & g_5 \\
1 & -1 & 0 & 0 & 0 & 0 \\
0 & 1 & -1 & 0 & 0 & 0 \\
0 & 0 & 1 & -1 & 0 & 0 \\
0 & 0 & 0 & 0 & 2 & -2 \\
\end{array}
\end{equation}
so that the relevant invariant residue forms are
\begin{equation}
\begin{array}{ccc}
& \beta & |\beta| \\
\hline
\omega_{(1^6)} & (1,1,1,1,1,1) & 1 \\ 
\omega_{(0^43^2)}  & (0,0,0,0,3,3) & 1 \\ 
\omega_{(2^6)} & (2,2,2,2,2,2) & 2 \\ 
\omega_{(1^44^2)} & (1,1,1,1,4,4) & 2 \\ 
\omega_{(3^40^2)} & (3,3,3,3,0,0) & 2 \\ 
\end{array}
\end{equation}

In this case we have
\begin{equation}
\begin{aligned}
\mathcal{Z}_{6^33} &= Span_\Z \left\{ Z((1^44^2),\beta')\oplus  Z((2^6),\beta')\right\} \\
%&=  Span_\Z \left\{\ -1^{\sum \beta_i'} \oplus \zeta_6^{2(\beta_0'+\beta_1'+µµ\beta_2'+\beta_3')+5(\beta_4'+\beta_5')} \right\}     
\end{aligned}
\end{equation}
By working these out for all $\beta'$, one finds that the generators of $\mathcal{Z}_{6^33}$ are $(1,0),(\zeta_6^2,0),(0,1)$, so that $\mathcal{Z}_{6^33} = A_2 \oplus A_1$ and $\mathcal{Z}_{6^33}^* = A_2^* \oplus A_1^*$.  

In order to determine $\Lambda_{phys}$, we now impose \eqref{eq:int_alg_cycle_flux} using $C_{0,0,0}$. As $C_{0,0,0} \cdot \omega_{(1^44^2)} = 0$  this only constrains the $A_1^*$ summand such thar
\begin{equation}\label{eq:extZ63z3_alg_cyc_cond}
\int_{C_{id}^{0,0,0}} G =\frac{2 \nu_{(2^6)}}{z_u z_{(2^6)}}  \int_{C_{id}^{0,0,0}} \omega_{(2^6)} = 
- 3^2 \nu_{(2^6)} \in \Z + \frac{1}{2}
\end{equation} 
As for the first example, we hence find
\begin{equation}
\nu_{(2^6)}=(2n_1+1)/2    
\end{equation}
together with $\bar{\nu}_{(1^44^2)}\in A_2^*$, i.e  
\begin{equation}
\bar{\nu}_{(1^44^2)} = n_2 + n_3 \frac{\zeta_{12}^{-1}}{\sqrt{3}}\, .
\end{equation}

The matrix $\widetilde{\rho_{IJ}}_{6^33}$ is:
\begin{equation}
    \widetilde{\rho_{IJ}}_{6^33}=\begin{pmatrix}
        \alpha_{(2^6)}&\alpha_{(3^40^2)}\\
        \alpha_{(3^40^2)}&0
    \end{pmatrix}
\end{equation}

The shortest $G \in \Lambda_{phys}$ for which $\rho$ has rank one is found for $n_1=n_2=n_3=0$ in which case we have 
\begin{equation}
\frac{\chi(X)}{24} -\frac{1}{2} G^2_{0} = 48 > 0 \,,
\end{equation}
The shortest $G$ with $\rho$ of full rank is found for $n_1=n_2=0$ and $n_3=1$, so that 
\begin{equation}
\frac{\chi(X)}{24} -\frac{1}{2} G^2_{0} = \frac{2610}{24} - \frac{1}{2}\left(\frac{3^5}{2}+ 3^2 2^6\right) = -240 < 0 
\end{equation}
where we have used $S_{(1^44^2)}= \frac{3^2}{2^6}$. There are hence no solutions with $\rho$ of maximal rank within the tadpole bound here.

\subsection{Symmetric under $\Gamma= \left(\Z/6\Z\right)^2 \times \left(\Z/3\Z\right) \times \left(\Z/2\Z\right)$}\label{sect:6^232}

Here the action of the group $\Gamma$ is given by 
\begin{equation}
\begin{array}{cccccc}
g_0 & g_1 & g_2 & g_3 & g_4 & g_5 \\
1 & -1 & 0 & 0 & 0 & 0 \\
0 & 1 & -1 & 0 & 0 & 0 \\
0 & 0 & 0 & 2 & -2 & 0 \\
0 & 0 & 0 & 0 & 3 & -3 \\
\end{array}
\end{equation}
and the invariant residue forms are:
\begin{equation}
\begin{array}{ccc}
& \beta & |\beta| \\
\hline
\omega_{(0^43^2)} & (0,0,0,0,3,3) & 1 \\ 
\omega_{(0^32^3)}  & (0,0,0,2,2,2) & 1 \\ 
\omega_{(0^341^2)}  & (0,0,0,4,1,1) & 1 \\ 
\omega_{(1^6)}  & (1,1,1,1,1,1) & 1 \\ 
\omega_{(1^330^2)}  & (1,1,1,3,0,0) & 1 \\ 
\omega_{(2^30^3)}  & (2,2,2,0,0,0) & 1 \\ 

\omega_{(0^34^3)} & (0,0,0,4,4,4) & 2 \\ 
\omega_{(1^44^2)} & (1,1,1,1,4,4) & 2 \\ 
\omega_{(1^33^3)} & (1,1,1,3,3,3) & 2 \\ 
\omega_{(2^303^2)} & (2,2,2,0,3,3) & 2 \\ 
\omega_{(2^6)} & (2,2,2,2,2,2) & 2 \\ 
\omega_{(2^341^2)} & (2,2,2,4,1,1) & 2 \\ 
\omega_{(3^31^3)} & (3,3,3,1,1,1) & 2 \\ 
\omega_{(3^40^2)} & (3,3,3,3,0,0) & 2 \\ 
\omega_{(4^30^3)} & (4,4,4,0,0,0) & 2 \\ 
\end{array}
\end{equation}

The fundamental difference compared to previous examples is that \eqref{eq:latticeZinclusion} is not an equality (when restricted to invariant forms) here, i.e. 
\begin{align*}
    \mathcal{Z}_{6^223}=Span_{\Z}\left\{ Z(\omega_{(0^34^3)},\beta'),Z(\omega_{(1^44^2)},\beta'),Z(\omega_{(1^33^3)},\beta'),Z(\omega_{(2^303^2)},\beta'),Z(\omega_{(2^6)},\beta')\right\}
\end{align*}
is not equal to $A_1 \oplus A_2^4$. A basis of the dual lattice $ \mathcal{Z}_{6^223}^*$ is given in matrix form by : 
\begin{equation*}
    P_{6^223}=
    \begin{pmatrix}
    \frac{i}{6\sqrt{3}}&-\frac{1}{4}+\frac{i}{12\sqrt{3}}&0&0&0&0&0&0&0\\
    0&0&-\frac{1}{2}+\frac{i}{2\sqrt{3}}&-\frac{3}{2}+\frac{i}{2\sqrt{3}}&0&0&0&0&0\\
    -\frac{3}{4}+\frac{i}{4\sqrt{3}}&-1+\frac{i}{2\sqrt{3}}&0&0&\frac{e^{i\pi\frac{5}{6}}}{\sqrt{3}}&-\frac{3}{2}+\frac{i}{2\sqrt{3}}&0&0&0\\
    0&0&0&0&0&0&\frac{e^{i\pi\frac{5}{6}}}{\sqrt{3}}&-\frac{3}{2}+\frac{i}{2\sqrt{3}}&0\\
   \frac{1}{6}&\frac{1}{3}&0&0&0&0&0&0&\frac{1}{2}\\
    \end{pmatrix}
\end{equation*}
and we can write a general element of $ \mathcal{Z}_{6^223}^*$ as
\begin{equation}\label{eq:lattice_basis_Zstar_6^223}
(\nu_{(0^34^3)},\nu_{(1^44^2)},\nu_{(1^33^3)},\nu_{(2^303^2)},\nu_{(2^6)}) = 
P_{6^223}\cdot \mu
\end{equation}
with $\mu \in \Z^9$. 
The Gram matrix of the lattice $ \mathcal{Z}_{6^223}^*$ is given by
\begin{align}\label{eq:gram6squared23}
    G_{6^223}&=\begin{pmatrix}987/2 & 555 & 0 & 0 & 240 & 672 & 0 & 0 & 81/2 \\ 
    555 & 1686 & 0 & 0 & 336 & 912 & 0 & 0 & 81 \\ 
    0 &  0 & 576 & 1440 & 0 & 0 & 0 & 0 & 0 \\
    0 &  0 &  1440 & 4032 & 0 & 0 & 0 & 0 & 0 \\
    240 &  336 &  0 &  0 & 192 & 480 & 0 & 0 & 0 \\
    672 &  912 &  0 &  0 &  480 & 1344 & 0 & 0 &0  \\
    0 &  0 &  0 &  0 &  0 &  0 & 216 & 540 & 0 \\
    0 &  0 &  0 &  0 &  0 &  0 &  540 & 1512 & 0 \\
    81/2 &  81 &  0 &  0 &  0 &  0 &  0 &  0 & 243/2\\  \end{pmatrix}
\end{align}
This is not an integral matrix as $\mathbb{Z}^*$ is not an integral lattice.

Finally, we have to impose G to lie in $\Lambda_{phys}$, we need to take care of the intersection of G with respect to a linear cycle. An appropriate choice here is again $C^{0,0,0}_{id}$, as before. While we have in total three 3-decomposable residues, namely $\omega_{(0^34^3)},\omega_{(1^33^3)},\omega_{(2^6)}$, we have fixed the permutation to be the identity, and thus only $\omega_{(2^6)}$ has a non-zero period with respect to this linear cycle since it is the only cycle which is 3-decomposable with respect to the identity permutation in this list.

Thus, to impose $G\in \Lambda_{phys}$, we are reduced to make sure that the period of the rescaled $\Tilde{\omega_{(2^6)}}$ is half integral with respect to $C^{0,0,0}_{id}$. For any given $\mu$, this condition becomes :
\begin{align}
    \frac{2\cdot (\frac{n_1}{6}+\frac{n_2}{3}+\frac{n_9}{2})}{z_u z_{(2^6)}}  \int_{C_{id}^{0,0,0}} \omega_{(2^6)} = 
- 3^2 \left(\frac{n_1}{6}+\frac{n_2}{3}+\frac{n_9}{2}\right) \in \Z +\frac{1}{2}
\end{align}
Note that $n_2$ has no impact on this condition, so we can freely choose it. The resulting constraint on $n_1$ and $n_9$ implies that $n_1 + 3 n_9$ is an odd integer, so that we can write
\begin{equation}
n_1 = 2k+1 - 3 n_9 
\end{equation}
for $k \in \Z$.

We are now ready to find all flux solutions for this model by generating all vectors in $ \mathcal{Z}_{6^223}^*$ up to some given length by computer, and then checking for each one if it is contained in $\Lambda_{phys}$. All lengths below $500$ appearing in $\Lambda_{phys}$ and the associated numbers of solutions are
\begin{center}
\begin{longtable}{|l|l|}
\hline \multicolumn{1}{|c|}{\textbf{Length}} & \multicolumn{1}{c|}{\textbf{Number of solutions}} \\ \hline 
\endfirsthead
243/2 & 2 \\
411/2 & 4 \\
603/2 & 4 \\
627/2 & 12 \\
675/2 & 12\\
843/2 & 24 \\
987/2 & 8 \\
\hline
\caption{Lengths in $\Lambda_{phys}$ below $500$.} \label{tab:results}
\end{longtable}
\end{center}

We can now work out 
\begin{equation}
\widetilde{\rho_{IJ}}_{6^223}=
\begin{pmatrix}
0&0&0&\alpha_{(3^40^2)}&0&\alpha_{(2^341^2)}\\
0&\alpha_{(4^30^3)}&\alpha_{(2^303^2)}&\alpha_{(3^31^3)}&0&\alpha_{(2^6)}\\
0&\alpha_{(2^303^2)}&0&0&0&0\\
\alpha_{(3^40^2)}&\alpha_{(3^31^3)}&0&0&\alpha_{(2^303^2)}&\alpha_{(1^33^3)}\\
0&0&0&\alpha_{(2^303^2)}&0&\alpha_{(1^44^2)}\\
\alpha_{(2^341^2)}&\alpha_{(2^6)}&0&\alpha_{(1^33^3)}&\alpha_{(1^44^2)}&\alpha_{(0^34^3)} \\
\end{pmatrix}
\end{equation}
for any of these solutions, and examine the relationship between its rank and $G^2$.

A first observation is that generically the same length can be associated to different ranks. Of course this can happen trivially if we rescale $G$, but also happens in a different manner here. For example, the length $G\cdot G=\frac{4083}{2}$ can correspond to $\widetilde{\rho_{IJ}}_{6^223}$ having rank $4$ or rank $6$, via for example the following solutions :
\begin{align*}
    \mu=&\begin{pmatrix}
        -1,&-1,&-1,&0,&0,&1,&0,&0,&0
    \end{pmatrix}\\
    \mu=&\begin{pmatrix}
        0,&-1,&0,&0,&-1,&1,&-1,&1,&-1
    \end{pmatrix}\\
\end{align*}

\begin{figure}[H]
\centering
\includegraphics[width=12cm]
{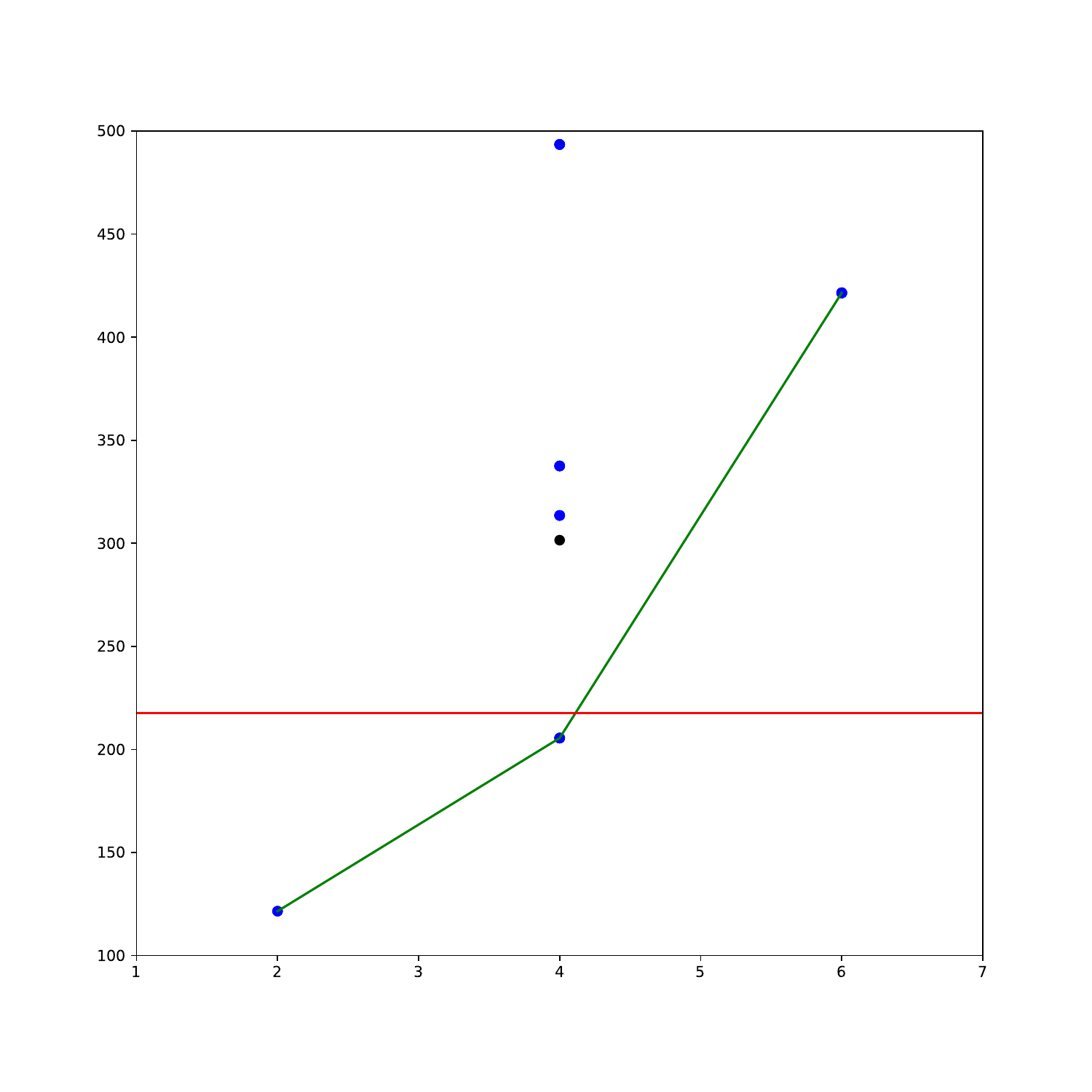}
\caption{A plot of the minimal lengths found for each rank of $\rho$. The horizontal axis shows the rank of $\rho$ and the vertical axis the tadpole contribution of the solutions. The red vertical line shows the tadpole bound.
\label{fig:fig1}}
\end{figure}

We have performed a scan over all lengths up to $1500$ and computed the associated rank of $\rho$ for all these solutions. This allows us to find the minimal length of $G$ for each rank of $\rho$. The result is shown in Table \ref{tab:minimas} and Figure \ref{fig:fig1}. 
\begin{center}
    \begin{table}[H]
        \centering
        \begin{tabular}{|c|c|}
        \hline
        \textbf{Rank} &\textbf{Minimum} \\
        \hline
         2 & 243/2\\
         4& 411/2 \\
         6 & 843/2\\
         \hline
        \end{tabular}
        \caption{Minimum for each rank}
        \label{tab:minimas}
    \end{table}
\end{center}
This is similar to the previous example, where the solutions found were only below the tadpole if the rank of the matrix was not full. The plot in Figure \ref{fig:fig1} shows the minimum lengths associated to every rank, as well as some further lattice points corresponding to rank four with non-minimal length. Interpolating the growth of $G^2$ with the rank of $\rho$ shows that the tadpole bound is crossed well before a maximal rank of $\rho$ is reached. 

For all lengths shown in Figure \ref{fig:fig1} except $\frac{603}{2}$, there exists a solution such that $\mu_i \in \{-1,0,1\} \forall i$. This is quite remarkable and indicates that constructing a basis of integral Hodge cycles using residues appears to be very efficient, at least when choosing appropriate linear combinations such as the ones in \eqref{eq:lattice_basis_Zstar_6^223}.

\subsection{Quotients}\label{sect:quotients}

The approach we have taken in this Section naturally lends itself to study flux solutions on quotients of the Fermat sextic by the groups $\Gamma$ considered. This appears to be a promising avenue to generate general Hodge cycles within the tadpole bound. On the one hand, the tadpole contribution is should be significantly smaller, as we expect the self-intersection number of a symmetric flux to be divided by the order of the group for the quotient. On the other hand, the tadpole contribution of the geometry should be equal to the Euler characteristic of a crepant resolution of the quotient, which is typically of a similar magnitude than the original fourfold.

Let us exemplify this for the simple case of the Fermat sextic and $\Gamma_{6^4}=(\Z/6\Z)^4$, where we can give a description using toric geometry. We first work out the Euler characteristic of a resolution. The family of sextic Calabi-Yau fourfolds is described as toric hypersurfaces by a pair of reflexive polytopes $\Delta,\Delta^*$ with vertices
\begin{equation}
\Delta^* = \begin{pmatrix}
 1 & 0 & 0 & 0 & 0 & -1\\
 0 & 1 & 0 & 0 & 0 & -1\\
 0 & 0 & 1 & 0 & 0 & -1\\
 0 & 0 & 0 & 1 & 0 & -1\\
 0 & 0 & 0 & 0 & 1 & -1\\
\end{pmatrix}\, , \hspace{1cm}
\Delta = \left(\begin{array}{rrrrrr}
-1 & -1 & -1 & -1 & -1 & 5 \\
-1 & -1 & -1 & -1 & 5 & -1 \\
-1 & -1 & -1 & 5 & -1 & -1 \\
-1 & -1 & 5 & -1 & -1 & -1 \\
-1 & 5 & -1 & -1 & -1 & -1
\end{array}\right)\, .
\end{equation}
Here $\Delta^*$ is the N-lattice polytope and $\Delta$ is the M-lattice polytope of the sextic fourfold $X$, and the mirror $X^\vee$ is found by reversing the roles of the two polytopes. Crucially, $X^\vee$ can also be found along the lines of \cite{Greene:1990ud} by taking (as resolution of) the quotient of $X/\Gamma_{6^4}$, and this is reflected in the face fan of $\Delta$ giving rise to the toric variety $\P^5/\Gamma_{6^4}$. It can be shown that $\Delta$ admits a fine and regular triangulation resulting in a projective crepant resolution $\widetilde{X_{\Gamma_{6^4}}}$ of $X/\Gamma_{6^4}$ with 
\begin{equation}
h^{1,1}(\widetilde{X_{\Gamma_{6^4}}}) = 426 \hspace{1cm} h^{3,1}(\widetilde{X_{\Gamma_{6^4}}}) = 1 \hspace{1cm} h^{2,1}(\widetilde{X_{\Gamma_{6^4}}}) = 0  
\end{equation}
so that $\chi(\widetilde{X_{\Gamma_{6^4}}}) =2610 = \chi(X)$ as expected for a mirror pair of Calabi-Yau fourfolds. 

We now work out the fate of the tadpole contribution of the flux. As $\omega_{(2^6)}$ is invariant under $\Gamma_{6^4}$ we will use the same notation to denote the image of this residue on the quotient. Following \cite{loyola2023toric}, we have that 
\begin{equation}
\frac{\int_X \omega_{(2^6)} \wedge \omega_{(2^6)}}{\int_{X/\Gamma_{6^4}}\omega_{(2^6)} \wedge \omega_{(2^6)}} = 
\frac{ \mbox{Vol}(\Delta) }{ \mbox{Vol}(\Delta^*) } = 
6^4 \, .
\end{equation}
where $\mbox{Vol}()$ is the lattice volume of the respective polytopes. The ratio here follows from the simple fact that the vertices of $\Delta^*$ span $N$, whereas the vertices of $\Delta$ span $N' \subset N$ with $N/N' = \Gamma_{6^4}$. The result above fits with the naive expectation that integrating an invariant form over a quotient is equal to the integral over the covering space divided by the order of the group. 

Similar results can be obtained for other groups $\Gamma$ as well. Here, the $N$-lattice polytope describing the quotient is given by a polytope which is a simplex with vertices $v_i$ satisfying $\sum v_i = 0$ such that $N/N' = \Gamma$, where $N'$ is again the sublattice of the $N$ lattice spanned by the $v_i$. It hence follows from the same argument as above that the tadpole contribution of the flux is reduced by $|\Gamma|$.

Given a flux symmetric under a finite group of symmetries, taking the quotient hence leads to a significant reduction of the tadpole contribution of the flux. This comes with another feature, however: the fourfolds $X/\Gamma$ are singular and the flux we have constructed is in general only defined on the singular fourfold $X/\Gamma$, i.e. these fluxes do not exist as properly quantized fluxes on a resolution of $X/\Gamma$. This is already indicated by the tadpole contribution of the flux being fractional in a way that does not originate from $c_2/2$. We expect this to be offset in the singular model by fractional M2 branes located at the orbifold singularities.

\section{Arithmetic and Obstructions}\label{sect:arithmetic}

Given a point in the complex structure moduli space of a Calabi-Yau fourfold $X$, the intersection product of primitive Hodge cycles naturally corresponds to an integral quadratic form $G$. The set of Hodge cycles below the Tadpole bound is then 
\begin{align*}
    S(G,T):=\{ k\: |\: \exists \, \mu \in \mathbb{Z}^{n}, \; G(\mu)=k\leq T \} \, ,
\end{align*}
with $T$ the associated tadpole bound and $n$ the dimension of $H^{2,2}\cap H^4(\Z)_{prim}$. While this set finite, performing an enumeration is computationally expensive and conceptually unsatisfactory, and we wish to find a necessary conditions for $S(G,T)$ to be non-empty. 

An integer $m$ is called representable by $G$ if there exists integers $\mu \in \Z^n$ s.t. $G(\mu) = m$. For an integer $m$ to be representable, we have to have corresponding representations of the p-adic reductions $m_p$ by $G_p$ and $\mu_p$ for every prime $p$:
\begin{align}\label{local_global}
    \mu\in \Z^n,\: m\in\Z,\: G(\mu)=m \implies \exists\, \mu_p\in \Z_p^n \, | \: G_p(\mu_p)=m_p \in \Z_p
\end{align}
This is a simple consequence of the fact that if an integral equation has an integral solution, then it also has a solution modulo every integer. Since every integer can be written uniquely as a product of prime powers, it suffices to limit ourselves to the p-adic integers and study solutions modulo every prime power.

Furthermore, the most relevant observation from the point of the view of the tadpole conjecture, is that there are typically obstructions for lengths to be represented by a given quadratic form. While it may only be possible to detect this by an exhaustive scan when working over $\Z$, it is often possible to quickly show that certain integers are not representable by studying the reduction modulo $p$.  

A standard example is the $A_2$ lattice, which is even, i.e. only even integers can be represented. Indeed, the associated quadratic form has a common factor of $2$ in every term. The consequence is that no odd numbers can be represented by the $A_2$ lattice because an odd number will reduce to $1$ modulo $2$, while the quadratic form of $A_2$ is identically $0$ modulo $2$, and thus there are no non-trivial solutions.

Let us exemplify this point of view for the problem treated in this work. Let us study the existence of solutions for the gram matrix $G(\Gamma_{6^232})$, \eqref{eq:gram6squared23}, i.e. the set $S(G(\Gamma_{6^232}),1500)$. A direct observation from the results shown in Table \ref{tab:results} is that there are only $7$ lengths in $\Lambda_{phys}$ below $500$, already well above the tadpole bound, indicating that the majority of integers below the tadpole bound cannot be represented.

In Table \ref{tab:all_length_sols}, we have generated a list of all integers up to $1500$ representable by $G(\Gamma_{6^232})$ which also includes their multiplicity. Note that this does not yet impose the physical quantization condition related to $c_2(X)/2$. 

As we can see right away, there are very few lengths represented by this quadratic form to begin with. In fact, there are only $108$ lengths represented up to $1500$. Furthermore, as the length increases, so does the number of solutions. 

Since it is valued in $\Z+\frac{1}{2}$ in general, we will multiply everything by a factor of 2. This allows us to recover the case of an integral quadratic form, and if we restrict ourselves to solutions that are odd, we recover $\Lambda_{phys}$ for $G(\Gamma_{6^232})$. This results in the following matrix :
\begin{align*}
    \Tilde{G}=\begin{pmatrix}987 & 1110 & 0 & 0 & 480 & 1344 & 0 & 0 & 81 \\ 
    1110 & 3372 & 0 & 0 & 672 & 1824 & 0 & 0 & 162 \\ 
    0 &  0 & 1152 & 2880 & 0 & 0 & 0 & 0 & 0 \\
    0 &  0 &  2880 & 8064 & 0 & 0 & 0 & 0 & 0 \\
    480 &  672 &  0 &  0 & 384 & 960 & 0 & 0 & 0 \\
    1344 &  1824 &  0 &  0 &  960 & 2688 & 0 & 0 & 0 \\
    0 &  0 &  0 &  0 &  0 &  0 & 432 & 1080 & 0 \\
    0 &  0 &  0 &  0 &  0 &  0 &  1080 & 3024 & 0 \\
    81 &  162 &  0 &  0 &  0 &  0 &  0 &  0 & 243\\  \end{pmatrix}
\end{align*}

As an example, let us study whether or not $433$ is represented by this matrix. Note that $433$ is odd and thus in principle can lie in $\Lambda_{phys}$. We know that if there is a solution $\mu\in\Z$, then there must be solutions in the p-adics $\mu_p \in \Z_p$ for every prime $p$. So we can first perform a reduction mod 3 of $\Tilde{G}$, which is identically $0$. Since $433=1\text{\,(mod 3)}$, we have an obstruction : there is no non-trivial solution in the 3-adics.

Let us also work out the obstruction modulo $2$ to see that it is indeed prime powers that matter, and not just the primes. For $n=1$, the quadratic form $\tilde{G}_2$ is:
\begin{align}
\tilde{G}_2  = n_1^2 + n_9^2
\end{align}
Since $433=1\text{\,(mod 2)}$ there are no obstructions. However, for $n=2$, the quadratic form $\tilde{G}_{2^2}$ is : 
\begin{align}\label{eq:mod4}
\tilde{G}_{2^2} =  3n_1^2 + 2n_1\cdot n_9 + 3n_9^2
\end{align}
Since $433=1\text{\,(mod 4)}$ there are in fact obstructions modulo $4$, as 
$\tilde{G}_{2^2}$ is always $0$ or $3$ modulo $4$.

While the existence of obstructions to solutions holds for quadratic forms in general, here it is quite important to note the algebro-geometric origin of the quadratic forms we are considering. From the geometric and physical context, one can hope to a priori determine the obstructions, which leaves to determine the representation of integers for which there are no obstructions, thus severely constraining the set of physical solutions. 

Finally, the key question we can ask then is does about the converse of \eqref{local_global}: if we have determined that there are no obstructions in the p-adics $\Z_p$, does this imply that we can lift the solutions in the p-adics $\Z_p$ to a solution in $\Z$ ? This is known as the local-global, or Hasse, principle \cite{cassels2008rational}. While it does not hold in general, given the specificity of this problem, with constraints coming from both geometry and physics, one can hope that studying this principle in this context might lead to new insights regarding the tadpole conjecture.

It would be interesting to combine the observation about quotients made above with the number theoretic approach outlined here. Dividing a given quadratic form by prime powers can be described in this language as removing obstructions to the representation of integers by this quadratic form. It would be very interesting to systematically investigate this further.

\section{Conclusions and Future Directions}
In this work we have continued the investigation of fluxed M-Theory compactifications on the Fermat sextic initiated in \cite{Braun:2011zm}. We have extended this work in two essential ways, from which can draw a number of lessons.

First, we have investigated fluxes defined via algebraic cycles that are not just of linear type, but also included two types of non-linear algebraic cycles: those of 'Aoki-Shioda' type and those of 'type 3'. This choice was motivated by these cycles spanning the entire $H^{2,2}(X) \cap H^4(X,\mathbb{Q})$. Looking for fluxes which stabilize all complex structure moduli, i.e. general Hodge cycles, we found solutions obeying the physical quantization condition with a tadpole contribution of $Q_{M2} = \tfrac12 G\cdot G = 587/4$. This is significantly less than the physically quantized general Hodge cycle constructed in \cite{Braun:2020jrx}, for which  $Q_{M2} = \tfrac12 G\cdot G = 775/4$. We hence find that the tadpole bound $\tfrac12 G\cdot G\leq 435/4$ is again violated by a sizeable margin. The ratio of the tadpole contribution of the flux and the number of stabilized moduli drops to $0.34$, which comes remarkably close to the ratio of $1/3$ originally conjectured in \cite{Bena:2020xrh}. 

The strength of this approach is that it is easy to implement the physical quantization condition, and straightforward to check the number of stabilized moduli by computing the resulting rank of $\rho$. However, it is far from clear if this approach can, in principle, be extended to describe any physically quantize flux as this requires the integral Hodge conjecture. Despite the fact that the set of algebraic cycles we are considering is enough to show the Hodge conjecture over $\Q$, we have shown that it fails to generate all of $H^{2,2} \cap H^4(X,\Z)$, i.e. even if the integral Hodge conjecture is true for the Fermat sextic, the types of algebraic cycles we are considering are still not enough to generate all integral Hodge cycles. 

The second extension of \cite{Braun:2011zm} we investigated was to study the physical quantization condition for Hodge cycles constructed from Griffiths residues. This guarantees to find all integral Hodge cycles and we can again, in principle, compute the rank of $\rho$ in a straightforward fashion. Due to the computational complexity we restricted ourselves to solutions with symmetries, which allowed us to study in examples how $G \cdot G$ grows with the rank of $\rho$. In all cases except taking the largest possible group of symmetries, $\Gamma= \left(\mathbb{Z}/6\mathbb{Z}\right)^4$, it was not possible to stabilize all (symmetric) moduli within the tadpole bound. 

Taking quotients significantly eases the tension between the desire to have a general Hodge cycle and remain within the tadpole bound. As remarked in Section \ref{sect:quotients}, this typically implies that the flux on the quotient is only defined on a singular fourfold. This is an interesting prospect in that singular fourfolds in M/F theory give rise to non-trivial gauge theory sectors. Apart from non-Higgsable clusters \cite{Morrison:2012np}, such sectors are absent at generic points in the moduli space, and it would be fascinating if the tadpole bound, together with the condition of having a general Hodge cycle, predicts the presence of gauge theory sectors in F-Theory. This idea resonates well with the results of \cite{Braun:2023pzd}. 

The two approaches we have taken rely on results about periods of algebraic cycles, and the construction of the integral middle cohomology as well as integrals of residues, respectively. It would be interesting to extend these results, and hence the present work, to other points in the moduli space of the sextic hypersurface, to hypersurfaces in weighted projective spaces or even toric varieties, or to complete intersections in products of projective spaces. Although the same tools used to establish the results we used for the Fermat sextic are in principle available in these cases, we still expect significant technical difficulties. 

The problem we were studying is ultimately arithmetic in nature, and one cannot help but wonder if the questions we are asking can more naturally be addressed in a number theoretic setting. As an example, we explained how the question which tadpoles can in principle be induced by the flux into the questions can be cast into the question of representability of integers by a quadratic form. The obstructions to numbers being representable can then be detected by studying this form in its p-adic reduction. It would be very interesting to push this further and study this approach in more generality. By fully exploiting the fact that the quadratic form one is studying is of geometric origin and is ultimately tied to the primes appearing in the defining equation of the fourfold in question, one can hope to make more far-reaching statements. In particular, it might be interesting to explore the implications of the existence of the elliptic fibration required to construct an F-Theory background. While this approach does not yet tell us what is the minimal length of a general Hodge cycle under appropriate quantization conditions, it nonetheless shows that the set of possible tadpoles can be restricted in a surprising fashion. 

The Fermat sextic studied here is known to be modular \cite{Schimmrigk:2008mp}, and it would be interesting to investigate the connection of the proposal of  \cite{Kachru:2020sio,Kachru:2020abh}, see also \cite{Candelas:2023yrg}, with the inner form between Hodge cycles. A recent study of modularity of Calabi-Yau fourfolds in relation to M-Theory flux vacua has appeared in \cite{Jockers:2023zzi}. The closely related arithmetic underlying solutions on quotients of $K3 \times K3$ of CM type was studied in \cite{Kanno:2017nub,Kanno:2020kxr}.

{\bf Acknowledgments:} The authors want to thank Hossein Movasati for the invitation to participate in his GADEPs Seminar, were this collaboration originally started. D.L.G. was supported  by grant \#2022/04705-8, S\~ao Paulo Research Foundation (FAPESP). R.V.L. was supported by Fondecyt ANID Postdoctoral Grant 3210020.

\appendix

\section{Lists of Linearly Independent Algebraic Cycles}\label{app:lin_indep_alg_cycles}
This appendix contains lists of linearly independent algebraic cycles of linear type, Aoki-Shioda type, and type 3. The tables list the powers of primitive roots of unity and permutation of homogeneous coordinates for each cycle.  

\begin{table}[!ht]
\tiny
\centering
\begin{tabular}{|l|l|l|}
		\hline
		Index&$(\ell_1,\ell_3,\ell_5)$&Permutations of linear cycles\\
		\hline
		\hline
		1,\ldots,14&(0,0,0)&$\Sigma\setminus$(0     1     2     3     4     5) \\
		\hline
		15,\ldots,299&(1,0,0),(2,0,0),(3,0,0),(4,0,0),(0,1,0)& $\Sigma$\\
		&(1,1,0),(2,1,0),(3,1,0),(4,1,0),(0,2,0)&\\
		&(1,2,0),(2,2,0),(3,2,0),(4,2,0),(0,3,0)&\\
		&(1,3,0),(2,3,0),(3,3,0),(4,3,0)&\\
		\hline
		300,\ldots,305&(0,4,0)&     
		(0     1     2     3     4     5),
		(0     1     2     4     3     5),
		(0     3     1     4     2     5),
		(0     1     2     5     3     4),
		(0     3     1     5     2     4),
		(0     4     1     5     2     3)\\
		\hline
		306,\ldots,308&(1,4,0)&      
		(0     3     1     4     2     5),
		(0     3     1     5     2     4),
		(0     4     1     5     2     3)\\
		\hline
		309,\ldots,311&(2,4,0)&      
		(0     3     1     4     2     5),
		(0     3     1     5     2     4),
		(0     4     1     5     2     3)\\
		\hline
		312,\ldots,314&(3,4,0)&
		(0     3     1     4     2     5),
		(0     3     1     5     2     4),
		(0     4     1     5     2     3)\\
		\hline
		315,\ldots,317&(4,4,0)&
		(0     3     1     4     2     5),
		(0     3     1     5     2     4),
		(0     4     1     5     2     3)\\
		\hline
		318,\ldots,617&
		(0,0,1),(1,0,1),(2,0,1),(3,0,1),(4,0,1)&$\Sigma$\\
		&(0,1,1),(1,1,1),(2,1,1),(3,1,1),(4,1,1)&\\
		&(0,2,1),(1,2,1),(2,2,1),(3,2,1),(4,2,1)&\\
		&(0,3,1),(1,3,1),(2,3,1),(3,3,1),(4,3,1)&\\
		\hline
		618,\ldots,623&(0,4,1)&     
		(0     1     2     3     4     5),
		(0     1     2     4     3     5),
		(0     3     1     4     2     5),
		(0     1     2     5     3     4),
		(0     3     1     5     2     4),
		(0     4     1     5     2     3)\\
		\hline
		624,\ldots,626&
		(1,4,1)&      
		(0     3     1     4     2     5),
		(0     3     1     5     2     4),
		(0     4     1     5     2     3)\\
		\hline
		627,\ldots,629&
		(2,4,1)&      
		(0     3     1     4     2     5),
		(0     3     1     5     2     4),
		(0     4     1     5     2     3)\\
		\hline
		630,\ldots,632&
		(3,4,1)&
		(0     3     1     4     2     5),
		(0     3     1     5     2     4),
		(0     4     1     5     2     3)\\
		\hline
		633,\ldots,635&
		(4,4,1)&
		(0     3     1     4     2     5),
		(0     3     1     5     2     4),
		(0     4     1     5     2     3)\\
		\hline
		636,\ldots,800&
		(0,0,2),(1,0,2),(2,0,2),(3,0,2),(4,0,2)&$\Sigma$\\
		&(0,1,2),(1,1,2),(2,1,2),(3,1,2),(4,1,2)&\\
		&(0,2,2)&\\
		\hline
		801,\ldots,814&
		(1,2,2)&$\Sigma\setminus$(0 1 2 5 3 4)\\
		\hline
		815,\ldots,828&
		(2,2,2)&$\Sigma\setminus$(0 1 2 5 3 4)\\
		\hline
		829,\ldots,842&
		(3,2,2)&$\Sigma\setminus$(0 1 2 5 3 4)\\
		\hline
		843,\ldots,856&
		(4,2,2)&$\Sigma\setminus$(0 1 2 5 3 4)\\
		\hline
		857,\ldots,871&
		(0,3,2)&$\Sigma$\\
		\hline
		872,\ldots,886&
		(1,3,2)&$\Sigma$\\
		\hline
		887,\ldots,901&
		(2,3,2)&$\Sigma$\\
		\hline
		902,\ldots,914&
		(3,3,2)&$\Sigma\setminus$(0 1 2 4 3 5), (0 4 1 2 3 5)\\
		\hline
		915,\ldots,927&
		(4,3,2)&$\Sigma\setminus$(0 1 2 4 3 5), (0 4 1 2 3 5)\\
		\hline
		928,\ldots,933&
		(0,4,2)&     
		(0     1     2     3     4     5),
		(0     1     2     4     3     5),
		(0     3     1     4     2     5),
		(0     1     2     5     3     4),
		(0     3     1     5     2     4),
		(0     4     1     5     2     3)\\
		\hline
		934&
		(1,4,2)& (0 3 1 4 2 5)\\
		\hline
		935&
		(2,4,2)& (0 3 1 4 2 5)\\
		\hline
		936&
		(3,4,2)& (0 3 1 4 2 5)\\
		\hline
		937&
		(4,4,2)& (0 3 1 4 2 5)\\
		\hline
		938,\ldots,948&
		(0,0,3)& 
		(0     1     2     3     4     5),
		(0     2     1     3     4     5),
		(0     3     1     2     4     5),
		(0     1     2     4     3     5),
		(0     2     1     4     3     5),
		(0     4     1     2     3     5),\\
		&&
		(0     3     1     4     2     5),
		(0     1     2     5     3     4),
		(0     2     1     5     3     4),
		(0     5     1     2     3     4),
		(0     4     1     5     2     3)\\
		\hline
		949,\ldots,956&
		(1,0,3)&
		(0     2     1     3     4     5),
		(0     3     1     2     4     5),
		(0     2     1     4     3     5),
		(0     4     1     2     3     5),
		(0     3     1     4     2     5),
		(0     2     1     5     3     4),\\&&
		(0     5     1     2     3     4),
		(0     4     1     5     2     3)\\
		\hline
		957,\ldots,964&
		(2,0,3)&
		(0     2     1     3     4     5),
		(0     3     1     2     4     5),
		(0     2     1     4     3     5),
		(0     4     1     2     3     5),
		(0     3     1     4     2     5),
		(0     2     1     5     3     4),\\
		&&
		(0     5     1     2     3     4),
		(0     4     1     5     2     3)\\
		\hline
		965,\ldots,970&
		(3,0,3)&
		(0     2     1     3     4     5),
		(0     3     1     2     4     5),
		(0     2     1     4     3     5),
		(0     3     1     4     2     5),
		(0     2     1     5     3     4),
		(0     5     1     2     3     4)\\
		\hline
		971,\ldots,976&
		(4,0,3)&
		(0     2     1     3     4     5),
		(0     3     1     2     4     5),
		(0     2     1     4     3     5),
		(0     3     1     4     2     5),
		(0     2     1     5     3     4),
		(0     5     1     2     3     4)\\
		\hline
		977,\ldots,984&
		(0,1,3)&
		(0     1     2     3     4     5),
		(0     2     1     3     4     5),
		(0     3     1     2     4     5),
		(0     2     1     4     3     5),
		(0     4     1     2     3     5),
		(0     3     1     4     2     5),\\
		&&
		(0     2     1     5     3     4),
		(0     5     1     2     3     4)\\
		\hline
		985,\ldots,989&
		(1,1,3)&
		(0     2     1     3     4     5),
		(0     3     1     2     4     5),
		(0     2     1     4     3     5),
		(0     4     1     2     3     5),
		(0     3     1     4     2     5)\\
		\hline
		990,\ldots,994&
		(2,1,3)&
		(0     2     1     3     4     5),
		(0     3     1     2     4     5),
		(0     2     1     4     3     5),
		(0     4     1     2     3     5),
		(0     3     1     4     2     5)\\
		\hline
		995&
		(3,1,3)& (0     2     1     3     4     5)\\
		\hline
		996&
		(4,1,3)& (0     2     1     3     4     5)\\
		\hline
		997,998,999&
		(0,2,3)& 
		(0     1     2     3     4     5)
		(0     2     1     3     4     5),
		(0     3     1     4     2     5)\\
		\hline
		1000,1001&
		(0,3,3)&
		(0     2     1     3     4     5),
		(0     3     1     2     4     5)\\
		\hline
\end{tabular}
\caption{List of linearly independent linear cycles. $\Sigma$ refers to the set of permutations \eqref{eq:sigma}.}
\label{table:linear cycles}
\end{table}

\newpage 

\begin{table}[h!]
\tiny
\centering
\begin{tabular}{|l|l|l|}

		\hline
		Index&$(\ell_0,\ell_2,\ell_3,\ell_5)$&Permutations of Aoki-Shioda cycles\\
		\hline
		\hline
		1002,\ldots,1337&(0,1,0,0)&$\Sigma'\setminus$
		(5     0     4     1     3     2),
		(5     1     4     0     3     2),
		(5     0     4     2     3     1),
		(5     2     4     0     3     1),\\
		&&
		(4     0     3     5     2     1),
		(5     0     3     4     2     1),
		(5     0     4     3     2     1),
		(4     3     5     0     2     1),\\
		&&
		(5     3     4     0     2     1),
		(5     1     4     2     3     0),
		(5     2     4     1     3     0),
		(4     1     3     5     2     0),\\
		&&
		(5     1     3     4     2     0),
		(5     1     4     3     2     0),
		(4     3     5     1     2     0),
		(5     3     4     1     2     0),\\
		&&
		(3     2     4     5     1     0),
		(4     2     3     5     1     0),
		(3     2     5     4     1     0),
		(5     2     3     4     1     0),\\
		&&
		(4     2     5     3     1     0),
		(5     2     4     3     1     0),
		(4     3     5     2     1     0),
		(5     3     4     2     1     0 )\\
		\hline
		1338,\ldots,1593&(1,1,0,0)& 
		( 0     1     2     3     4     5),(
		1     0     2     3     4     5),(
		2     0     1     3     4     5),(
		0     1     3     2     4     5),(
		1     0     3     2     4     5),(
		3     0     1     2     4     5),\\&&(
		0     2     3     1     4     5),(
		2     0     3     1     4     5),(
		3     0     2     1     4     5),(
		1     2     3     0     4     5),(
		2     1     3     0     4     5),(
		3     1     2     0     4     5),\\&&(
		0     1     2     4     3     5),(
		1     0     2     4     3     5),(
		2     0     1     4     3     5),(
		0     1     4     2     3     5),(
		1     0     4     2     3     5),(
		4     0     1     2     3     5),\\&&(
		0     2     4     1     3     5),(
		2     0     4     1     3     5),(
		4     0     2     1     3     5),(
		1     2     4     0     3     5),(
		2     1     4     0     3     5),(
		4     1     2     0     3     5),\\&&(
		0     1     3     4     2     5),(
		1     0     3     4     2     5),(
		3     0     1     4     2     5),(
		0     1     4     3     2     5),(
		1     0     4     3     2     5),(
		4     0     1     3     2     5),\\&&(
		0     3     4     1     2     5),(
		3     0     4     1     2     5),(
		4     0     3     1     2     5),(
		1     3     4     0     2     5),(
		3     1     4     0     2     5),(
		4     1     3     0     2     5),\\&&(
		0     2     3     4     1     5),(
		2     0     3     4     1     5),(
		3     0     2     4     1     5),(
		0     2     4     3     1     5),(
		2     0     4     3     1     5),(
		4     0     2     3     1     5),\\&&(
		0     3     4     2     1     5),(
		3     0     4     2     1     5),(
		4     0     3     2     1     5),(
		2     3     4     0     1     5),(
		3     2     4     0     1     5),(
		4     2     3     0     1     5),\\&&(
		1     2     3     4     0     5),(
		2     1     3     4     0     5),(
		3     1     2     4     0     5),(
		1     2     4     3     0     5),(
		2     1     4     3     0     5),(
		4     1     2     3     0     5),\\&&(
		1     3     4     2     0     5),(
		3     1     4     2     0     5),(
		4     1     3     2     0     5),(
		2     3     4     1     0     5),(
		3     2     4     1     0     5),(
		4     2     3     1     0     5),\\&&(
		0     1     2     3     5     4),(
		1     0     2     3     5     4),(
		2     0     1     3     5     4),(
		0     1     3     2     5     4),(
		1     0     3     2     5     4),(
		3     0     1     2     5     4),\\&&(
		0     2     3     1     5     4),(
		2     0     3     1     5     4),(
		3     0     2     1     5     4),(
		1     2     3     0     5     4),(
		2     1     3     0     5     4),(
		3     1     2     0     5     4),\\&&(
		0     1     2     5     3     4),(
		1     0     2     5     3     4),(
		2     0     1     5     3     4),(
		0     1     5     2     3     4),(
		1     0     5     2     3     4),(
		5     0     1     2     3     4),\\&&(
		0     2     5     1     3     4),(
		2     0     5     1     3     4),(
		5     0     2     1     3     4),(
		1     2     5     0     3     4),(
		2     1     5     0     3     4),(
		5     1     2     0     3     4),\\&&(
		0     1     3     5     2     4),(
		1     0     3     5     2     4),(
		3     0     1     5     2     4),(
		0     1     5     3     2     4),(
		1     0     5     3     2     4),(
		5     0     1     3     2     4),\\&&(
		0     3     5     1     2     4),(
		3     0     5     1     2     4),(
		5     0     3     1     2     4),(
		1     3     5     0     2     4),(
		3     1     5     0     2     4),(
		5     1     3     0     2     4),\\&&(
		0     2     3     5     1     4),(
		2     0     3     5     1     4),(
		3     0     2     5     1     4),(
		0     2     5     3     1     4),(
		2     0     5     3     1     4),(
		5     0     2     3     1     4),\\&&(
		0     3     5     2     1     4),(
		3     0     5     2     1     4),(
		5     0     3     2     1     4),(
		2     3     5     0     1     4),(
		3     2     5     0     1     4),(
		5     2     3     0     1     4),\\&&(
		1     2     3     5     0     4),(
		2     1     3     5     0     4),(
		3     1     2     5     0     4),(
		1     2     5     3     0     4),(
		2     1     5     3     0     4),(
		5     1     2     3     0     4),\\&&(
		1     3     5     2     0     4),(
		3     1     5     2     0     4),(
		5     1     3     2     0     4),(
		2     3     5     1     0     4),(
		5     2     3     1     0     4),(
		0     1     2     4     5     3),\\&&(
		1     0     2     4     5     3),(
		2     0     1     4     5     3),(
		0     1     4     2     5     3),(
		1     0     4     2     5     3),(
		4     0     1     2     5     3),(
		0     2     4     1     5     3),\\&&(
		2     0     4     1     5     3),(
		4     0     2     1     5     3),(
		1     2     4     0     5     3),(
		2     1     4     0     5     3),(
		4     1     2     0     5     3),(
		0     1     2     5     4     3),\\&&(
		1     0     2     5     4     3),(
		2     0     1     5     4     3),(
		0     1     5     2     4     3),(
		1     0     5     2     4     3),(
		5     0     1     2     4     3),(
		0     2     5     1     4     3),\\&&(
		2     0     5     1     4     3),(
		1     2     5     0     4     3),(
		0     1     4     5     2     3),(
		1     0     4     5     2     3),(
		4     0     1     5     2     3),(
		0     1     5     4     2     3),\\&&(
		1     0     5     4     2     3),(
		5     0     1     4     2     3),(
		0     4     5     1     2     3),(
		4     0     5     1     2     3),(
		5     0     4     1     2     3),(
		1     4     5     0     2     3),\\&&(
		4     1     5     0     2     3),(
		5     1     4     0     2     3),(
		0     2     4     5     1     3),(
		2     0     4     5     1     3),(
		4     0     2     5     1     3),(
		0     2     5     4     1     3),\\&&(
		2     0     5     4     1     3),(
		5     0     2     4     1     3),(
		0     4     5     2     1     3),(
		4     0     5     2     1     3),(
		2     4     5     0     1     3),(
		4     2     5     0     1     3),\\&&(
		5     2     4     0     1     3),(
		1     2     4     5     0     3),(
		2     1     4     5     0     3),(
		4     1     2     5     0     3),(
		1     2     5     4     0     3),(
		5     1     2     4     0     3),\\&&(
		1     4     5     2     0     3),(
		4     1     5     2     0     3),(
		2     4     5     1     0     3),(
		0     1     3     4     5     2),(
		1     0     3     4     5     2),(
		3     0     1     4     5     2),\\&&(
		0     1     4     3     5     2),(
		1     0     4     3     5     2),(
		4     0     1     3     5     2),(
		0     3     4     1     5     2),(
		3     0     4     1     5     2),(
		4     0     3     1     5     2),\\&&(
		1     3     4     0     5     2),(
		3     1     4     0     5     2),(
		4     1     3     0     5     2),(
		0     1     3     5     4     2),(
		1     0     3     5     4     2),(
		3     0     1     5     4     2),\\&&(
		0     1     5     3     4     2),(
		1     0     5     3     4     2),(
		5     0     1     3     4     2),(
		0     3     5     1     4     2),(
		3     0     5     1     4     2),(
		1     3     5     0     4     2),\\&&(
		0     1     4     5     3     2),(
		1     0     4     5     3     2),(
		4     0     1     5     3     2),(
		0     1     5     4     3     2),(
		4     0     5     1     3     2),(
		0     3     4     5     1     2),\\&&(
		3     0     4     5     1     2),(
		0     3     5     4     1     2),(
		3     0     5     4     1     2),(
		0     4     5     3     1     2),(
		4     0     5     3     1     2),(
		3     4     5     0     1     2),\\&&(
		4     3     5     0     1     2),(
		5     3     4     0     1     2),(
		1     3     4     5     0     2),(
		3     1     4     5     0     2),(
		1     3     5     4     0     2),(
		1     4     5     3     0     2),\\&&(
		3     4     5     1     0     2),(
		0     2     3     4     5     1),(
		2     0     3     4     5     1),(
		3     0     2     4     5     1),(
		0     2     4     3     5     1),(
		2     0     4     3     5     1),\\&&(
		4     0     2     3     5     1),(
		0     3     4     2     5     1),(
		3     0     4     2     5     1),(
		2     3     4     0     5     1),(
		3     2     4     0     5     1),(
		4     2     3     0     5     1),\\&&(
		0     2     3     5     4     1),(
		2     0     3     5     4     1),(
		0     2     5     3     4     1),(
		2     0     5     3     4     1),(
		5     0     2     3     4     1),(
		3     0     5     2     4     1),\\&&(
		2     3     5     0     4     1),(
		0     2     4     5     3     1),(
		2     0     4     5     3     1),(
		4     0     5     2     3     1),(
		0     3     4     5     2     1),(
		3     0     4     5     2     1),\\&&(
		2     3     4     5     0     1),(
		2     3     5     4     0     1),(
		2     4     5     3     0     1),(
		3     4     5     2     0     1),(
		1     2     3     4     5     0),(
		2     1     3     4     5     0),\\&&(
		3     1     2     4     5     0),(
		1     2     4     3     5     0),(
		2     1     4     3     5     0),(
		4     1     2     3     5     0),(
		1     3     4     2     5     0),(
		3     1     4     2     5     0),\\&&(
		2     3     4     1     5     0),(
		1     2     3     5     4     0),(
		1     2     5     3     4     0),(
		2     1     5     3     4     0),(
		5     1     2     3     4     0),(
		3     1     5     2     4     0),\\&&(
		1     2     4     5     3     0),(
		4     1     5     2     3     0),(
		1     3     4     5     2     0),(
		2     3     4     5     1     0)
		\\
		\hline
		1594,\ldots,1645&(0,0,0,1)& 
		( 0     1     2     3     4     5),(
		1     0     2     3     4     5),(
		0     1     3     2     4     5),(
		1     0     3     2     4     5),(
		0     2     3     1     4     5),(
		2     0     3     1     4     5),\\&&(
		1     2     3     0     4     5),(
		2     1     3     0     4     5),(
		0     1     2     4     3     5),(
		1     0     2     4     3     5),(
		0     1     4     2     3     5),(
		1     0     4     2     3     5),\\&&(
		0     2     4     1     3     5),(
		2     0     4     1     3     5),(
		1     2     4     0     3     5),(
		0     1     3     4     2     5),(
		1     0     3     4     2     5),(
		0     1     4     3     2     5),\\&&(
		3     0     4     1     2     5),(
		2     0     3     4     1     5),(
		0     1     2     3     5     4),(
		1     0     2     3     5     4),(
		0     1     3     2     5     4),(
		1     0     3     2     5     4),\\&&(
		2     0     3     1     5     4),(
		0     1     2     5     3     4),(
		1     0     2     5     3     4),(
		0     1     5     2     3     4),(
		1     0     5     2     3     4),(
		0     2     5     1     3     4),\\&&(
		2     0     5     1     3     4),(
		1     2     5     0     3     4),(
		0     1     3     5     2     4),(
		1     0     3     5     2     4),(
		0     1     5     3     2     4),(
		3     0     5     1     2     4),\\&&(
		2     0     3     5     1     4),(
		0     1     2     4     5     3),(
		1     0     2     4     5     3),(
		0     1     4     2     5     3),(
		1     0     4     2     5     3),(
		2     0     4     1     5     3),\\&&(
		0     1     4     5     2     3),(
		1     0     4     5     2     3),(
		0     1     5     4     2     3),(
		4     0     5     1     2     3),(
		2     0     4     5     1     3),(
		0     1     3     4     5     2),\\&&(
		1     0     3     4     5     2),(
		3     0     4     1     5     2),(
		3     0     4     5     1     2),(
		2     0     3     4     5     1)\\
		\hline
		1646,\ldots, 1687 &(1,0,0,1)&
		( 0     1     2     3     4     5),(
		1     0     2     3     4     5),(
		0     1     3     2     4     5),(
		1     0     3     2     4     5),(
		0     2     3     1     4     5),(
		2     0     3     1     4     5),\\&&(
		1     2     3     0     4     5),(
		2     1     3     0     4     5),(
		0     1     2     4     3     5),(
		1     0     2     4     3     5),(
		0     1     4     2     3     5),(
		1     0     4     2     3     5),\\&&(
		0     2     4     1     3     5),(
		2     0     4     1     3     5),(
		1     2     4     0     3     5),(
		0     1     3     4     2     5),(
		1     0     3     4     2     5),(
		0     1     4     3     2     5),\\&&(
		3     0     4     1     2     5),(
		2     0     3     4     1     5),(
		0     1     2     3     5     4),(
		1     0     2     3     5     4),(
		0     1     3     2     5     4),(
		0     1     2     5     3     4),\\&&(
		1     0     2     5     3     4),(
		0     1     5     2     3     4),(
		1     0     5     2     3     4),(
		0     2     5     1     3     4),(
		2     0     5     1     3     4),(
		1     2     5     0     3     4),\\&&(
		0     1     3     5     2     4),(
		0     1     5     3     2     4),(
		3     0     5     1     2     4),(
		0     1     2     4     5     3),(
		1     0     2     4     5     3),(
		0     1     4     2     5     3),\\&&(
		0     1     4     5     2     3),(
		0     1     5     4     2     3),(
		4     0     5     1     2     3),(
		0     1     3     4     5     2),(
		1     0     3     4     5     2),(
		2     0     3     4     5     1)\\
		\hline
		1688,\ldots,1703& (0,1,0,1)&
		( 0     1     2     3     4     5),(
		0     1     3     2     4     5),(
		0     2     3     1     4     5),(
		1     2     3     0     4     5),(
		0     1     2     4     3     5),(
		0     1     4     2     3     5),\\&&(
		0     1     3     4     2     5),(
		0     2     3     4     1     5),(
		1     2     3     4     0     5),(
		0     1     2     3     5     4),(
		0     1     2     5     3     4),(
		0     1     5     2     3     4),\\&&(
		0     1     2     4     5     3),(
		0     1     3     4     5     2),(
		0     2     3     4     5     1),(
		1     2     3     4     5     0)\\
		\hline
		1704,\ldots,1719&(1,1,0,1)&
		( 0     1     2     3     4     5),(
		0     1     3     2     4     5),(
		0     2     3     1     4     5),(
		1     2     3     0     4     5),(
		0     1     2     4     3     5),(
		0     1     4     2     3     5),\\&&(
		0     1     3     4     2     5),(
		0     2     3     4     1     5),(
		1     2     3     4     0     5),(
		0     1     2     3     5     4),(
		0     1     2     5     3     4),(
		0     1     5     2     3     4),\\&&(
		0     1     2     4     5     3),(
		0     1     3     4     5     2),(
		0     2     3     4     5     1),(
		1     2     3     4     5     0)\\
		\hline
		1720&(0,1,0,2)&(0,1,2,3,4,5)\\
		\hline
		1721&(1,1,0,2)&(0,1,2,3,4,5)\\
		\hline		
\end{tabular}
\caption{List of linearly independent Aoki-Shioda cycles. $\Sigma'$ refers to the set of permutations with $\sigma(1) < \sigma(2)$.}
\label{table:AS cycles}
\end{table}

\begin{table}[H]
\tiny
\centering
\begin{tabular}{|l|l|l|}
		\hline
		Index& $(\ell_1,\ell_2,\ell_3,\ell_4,\ell_5)$&Permutations of type 3 cycles\\
		\hline
		\hline
		1722,\ldots,1730&(0,0,0,0,0)&     
		( 0     1     2     3     4     5),(
		1     0     2     3     4     5),(
		0     1     3     2     4     5),(
		0     1     2     4     3     5),(
		1     0     2     4     3     5),(
		0     1     4     2     3     5),\\&&(
		0     1     3     4     2     5),(
		1     0     3     4     2     5),(
		2     0     3     4     1     5)\\
		\hline
		1731,\ldots,1744&(1,0,0,0,0)&
		( 0     1     2     3     4     5),(
		1     0     2     3     4     5),(
		2     0     1     3     4     5),(
		0     1     3     2     4     5),(
		0     1     2     4     3     5),(
		1     0     2     4     3     5),\\&&(
		2     0     1     4     3     5),(
		0     1     4     2     3     5),(
		0     1     3     4     2     5),(
		1     0     3     4     2     5),(
		3     0     1     4     2     5),(
		0     2     3     4     1     5),\\&&(
		2     0     3     4     1     5),(
		1     2     3     4     0     5)\\
		\hline
		1745,\ldots,1749&(3,0,0,0,0)&
		( 0     1     2     3     4     5),(
		0     1     2     4     3     5),(
		0     1     3     4     2     5),(
		0     2     3     4     1     5),(
		1     2     3     4     0     5)	     \\
		\hline
		1750&(1,0,0,1,0)&( 0 1 2 3 4 5)\\
		\hline
		1751&(3,0,0,1,0)&( 0 1 2 3 4 5)\\
		\hline
\end{tabular}
\caption{List of linearly independent type 3 cycles.}
\label{table:t3 cycles}
\end{table}

\newpage

\section{Subgroups of $(\Z/6\Z)^4$ acting on the Fermat Sextic}\label{app:subgroupsZ64}

Below is a list of the occuring dimensions of invariant subspaces $H^{3,1}(X)_{inv} \subset H^{3,1}(X)$ and $H^{2,2}(X)_{inv} \subset H^{2,2}(X)$ 
for all subgroups of $(\Z/6\Z)^4$. Note that a single entry potentially corresponds to genuinely different subgroups of $(\Z/6\Z)^4$, i.e. 
subgroups which are not identified by permuting the $x_i$. As can be seen from the arrangement of the table, there is a matching between 
cases with $|\Gamma|=k$ and cases with $|\Gamma^\vee|=6^4/k$, which is a consequence of mirror symmetry. 
\begin{longtable}[H]{ccc||ccc}
\caption{Orders and dimensions of invariant subspaces for all subgroups of 
$(\Z/6\Z)^4$.}\label{tab:possible_groups} 
\endfirsthead
\endhead

$h^{3,1}_{inv}(X)$ & $h^{2,2}_{inv}(X)$ & $|\Gamma|$ & 
$h^{3,1}_{inv}(X)$ & $h^{2,2}_{inv}(X)$ & $|\Gamma|$\\
\hline
1& 426& 1751& 1296& 1& 1\\
 2& 226& 903& 648& 2& 3\\
 3& 138& 563& 432& 1& 7\\
 3& 144& 587& 432& 3& 3\\
 3& 162& 611& 432& 3& 5\\
 4& 126& 479& 324& 4& 7\\
 6& 70& 291& 216& 2& 11\\
 6& 72& 315& 216& 4& 9\\
 6& 74& 291& 216& 4& 11\\
 6& 76& 303& 216& 5& 7\\
 6& 84& 309& 216& 6& 7\\
 6& 86& 315& 216& 6& 9\\
 8& 76& 267& 162& 8& 15\\
 9& 48& 191& 144& 3& 23\\
 9& 52& 191& 144& 5& 17\\
 9& 62& 215& 144& 7& 17\\
 9& 66& 215& 144& 9& 11\\
 12& 36& 155& 108& 4& 19\\
 12& 36& 179& 108& 6& 19\\
 12& 38& 155& 108& 6& 21\\
 12& 40& 155& 108& 6& 23\\
 12& 40& 167& 108& 8& 17\\
 12& 42& 155& 108& 8& 19\\
 12& 42& 161& 108& 8& 21\\
 12& 46& 161& 108& 10& 15\\
 12& 48& 167& 108& 10& 19\\
 12& 60& 149& 108& 12& 17\\
 16& 51& 161& 81& 16& 31\\
 18& 24& 99& 72& 6& 33\\
 18& 24& 103& 72& 6& 35\\
 18& 26& 97& 72& 7& 29\\
 18& 28& 99& 72& 8& 29\\
 18& 30& 111& 72& 8& 33\\
 18& 30& 119& 72& 10& 27\\
 18& 32& 109& 72& 10& 31\\
 18& 34& 107& 72& 11& 29\\
 18& 34& 111& 72& 12& 25\\
 18& 36& 95& 72& 12& 27\\
 18& 42& 99& 72& 12& 31\\
 18& 42& 103& 72& 14& 29\\
 24& 22& 87& 54& 10& 35\\
 24& 22& 99& 54& 12& 43\\
 24& 24& 87& 54& 14& 35\\
 24& 24& 93& 54& 14& 37\\
 24& 27& 87& 54& 14& 43\\
 24& 34& 81& 54& 16& 35\\
 24& 36& 87& 54& 18& 39\\
 27& 22& 67& 48& 15& 53\\
 27& 28& 75& 48& 15& 59\\
 27& 30& 83& 48& 21& 47\\
 36& 12& 53& 36& 12& 53\\
 36& 12& 57& 36& 12& 55\\
 36& 14& 51& 36& 12& 57\\
 36& 14& 63& 36& 12& 59\\
 36& 16& 57& 36& 14& 51\\
 36& 18& 49& 36& 14& 55\\
 36& 18& 55& 36& 14& 63\\
 36& 20& 51& 36& 16& 55\\
 36& 22& 49& 36& 16& 57\\
 36& 22& 53& 36& 16& 61\\
 36& 30& 41& 36& 18& 49 
\end{longtable}

\section{Short lattice points for $\Gamma= \left(\Z/6\Z\right)^2 \times \left(\Z/3\Z\right) \times \left(\Z/2\Z\right)$}

This is a table of shortest lengths appearing in the lattice $\mathcal{Z}^*_{6^232}$ for the group $\Gamma= \left(\Z/6\Z\right)^2 \times \left(\Z/3\Z\right) \times \left(\Z/2\Z\right)s$ discussed in Section \ref{sect:6^232}. 

\begin{center}
\begin{longtable}{|l|l|l|l|}
\caption{Lengths below 1500 and number of lattice points for $\mathcal{Z}^*_{6^232}$.} \label{tab:all_length_sols} \\

\hline \multicolumn{1}{|c|}{\textbf{Length}} & \multicolumn{1}{c|}{\textbf{Number of solutions}} & \multicolumn{1}{|c|}{\textbf{Length}}& \multicolumn{1}{c|}{\textbf{Number of solutions}} \\ \hline 
\endfirsthead

\multicolumn{4}{c}%
{{\bfseries \tablename\, \thetable{} -- continued from previous page}} \\

\hline \multicolumn{1}{|c|}{\textbf{Length}} & \multicolumn{1}{c|}{\textbf{Number of solutions}} & \multicolumn{1}{|c|}{\textbf{Length}}& \multicolumn{1}{c|}{\textbf{Number of solutions}} \\ \hline 
\endhead

\hline \multicolumn{4}{|r|}{{Continued on next page}} \\ \hline
\endfoot

\hline \hline
\endlastfoot

    0 & 1 & 1971/2 & 12 \\
243/2 & 2 & 990 & 24 \\
192 & 6 & 1995/2 & 264 \\
411/2 & 4 & 1008 & 36 \\
216 & 6 & 2067/2 & 48 \\
246 & 4 & 1038 & 264 \\
603/2 & 4 & 1056 & 64 \\
627/2 & 12 & 1062 & 24 \\
675/2 & 12 & 2139/2 & 112 \\
342 & 4 & 1080 & 24 \\
408 & 40 & 2187/2 & 146 \\
843/2 & 24 & 2211/2 & 548 \\
462 & 24 & 1110 & 88 \\
486 & 2 & 2259/2 & 36 \\
987/2 & 8 & 1134 & 156 \\
504 & 4 & 2283/2 & 56 \\
1035/2 & 24 & 1152 & 64 \\
1059/2 & 76 & 2331/2 & 112 \\
534 & 8 & 2355/2 & 96 \\
558 & 24 & 1182 & 48 \\
576 & 12 & 1200 & 264 \\
1179/2 & 6 & 2403/2 & 24 \\
624 & 24 & 1206 & 88 \\
1251/2 & 4 & 1224 & 72 \\
630 & 6 & 2475/2 & 42 \\
648 & 6 & 1254 & 84 \\
678 & 12 & 1272 & 112 \\
696 & 8 & 2547/2 & 96 \\
1395/2 & 24 & 1278 & 180 \\
702 & 12 & 2571/2 & 396 \\
1419/2 & 48 & 1296 & 144 \\
720 & 24 & 2619/2 & 12 \\
1491/2 & 24 & 2643/2 & 264 \\
750 & 48 & 1326 & 456 \\
768 & 42 & 1344 & 116 \\
1539/2 & 12 & 2691/2 & 144 \\
1563/2 & 44 & 1350 & 12 \\
792 & 78 & 2715/2 & 248 \\
1611/2 & 36 & 1368 & 328 \\
1635/2 & 8 & 2763/2 & 384 \\
822 & 44 & 2787/2 & 88 \\
840 & 36 & 1398 & 200 \\
1683/2 & 24 & 1416 & 252 \\
846 & 36 & 2835/2 & 144 \\
1707/2 & 28 & 1422 & 384 \\
864 & 6 & 2859/2 & 308 \\
1755/2 & 24 & 1440 & 126 \\
1779/2 & 84 & 2907/2 & 84 \\
894 & 96 & 2931/2 & 168 \\
912 & 48 & 1470 & 768 \\
1827/2 & 150 & 1488 & 384 \\
918 & 24 & 2979/2 & 520 \\
1899/2 & 28 & 1494 & 48 
\end{longtable}
\end{center}

\clearpage

\providecommand{\href}[2]{#2}\begingroup\raggedright\endgroup

\end{document}